\documentclass[10pt, conference, letterpaper]{IEEEtran}
\IEEEoverridecommandlockouts
\usepackage{cite}
\usepackage{color}
\usepackage{amsmath,amssymb,amsfonts}
\usepackage{algorithmic}
\usepackage{graphicx}
\usepackage{textcomp}
\def\BibTeX{{\rm B\kern-.05em{\sc i\kern-.025em b}\kern-.08em
    T\kern-.1667em\lower.7ex\hbox{E}\kern-.125emX}}

\usepackage[ruled]{algorithm2e}

\usepackage{url}

\def\BibTeX{{\rm B\kern-.05em{\sc i\kern-.025em b}\kern-.08em T\kern-.1667em\lower.7ex\hbox{E}\kern-.125emX}}

\newcommand {\mymarginpar}[1]{\marginpar{#1}}
\renewcommand {\marginpar}[1]{}

\def\_{\rule{.3em}{.15ex}}      

\newcommand{\ls}[1]
   {\dimen0=\fontdimen6\the\font
    \lineskip=#1\dimen0
    \advance\lineskip.5\fontdimen5\the\font
    \advance\lineskip-\dimen0
    \lineskiplimit=.9\lineskip
    \baselineskip=\lineskip
    \advance\baselineskip\dimen0
    \normallineskip\lineskip
    \normallineskiplimit\lineskiplimit
    \normalbaselineskip\baselineskip
    \ignorespaces
   }

\newcommand {\bearn}{\begin{eqnarray*}}
\newcommand {\eearn}{\end{eqnarray*}}
\newcommand {\barr}{\begin{array}}
\newcommand {\earr}{\end{array}}

\newcommand {\N}{{\cal N}}

\newtheorem{definition}{Definition}
\newtheorem{property}[definition]{Property}
\newtheorem{proposition}[definition]{Proposition}
\newtheorem{lemma}[definition]{Lemma}
\newtheorem{theorem}[definition]{Theorem}
\newtheorem{corollary}[definition]{Corollary}
\newtheorem{example}{Example}
\newtheorem{remark}[definition]{Remark}

\newcommand {\benum} {\begin{enumerate}}
\newcommand {\eenum} {\end{enumerate}}

\newcommand {\bdesc} {\begin{description}}
\newcommand {\edesc} {\end{description}}

\newcommand {\bfig}[2] {\begin{figure}
  \centering
  \includegraphics[width=#2]{#1}}
\newcommand {\brotatefig}[2] {\begin{figure}[htbp]
                        \centerline {
                         \epsfig{figure={#1},clip=,angle=-90,width={#2}}}}
\newcommand {\bfigfirst}[2] {\begin{figure}[h]
                        \centerline {
                        \setlength{\epsfxsize}{#2}
                        \epsffile{#1}}}
\newcommand {\efig}[2]{ \caption{#2}
                        \label{fig:#1}
                        \end{figure}
                        \mymarginpar{fig:#1}}
\newcommand {\erotatefig}[2]{ \caption{#2}
                        \label{fig:#1}
                        \end{figure}
                        \mymarginpar{fig:#1}}
\newcommand {\rfig}[1]{Figure \ref{fig:#1}}

\newcommand {\btab}[1]{
                       \begin{table}
                       \centering
                       \begin{tabular}{#1}}
\newcommand {\etab}[3] {
                       \end{tabular}
                       \caption[#3]{#2}
                       \label{tab:#1}
                       \end{table}
                       \mymarginpar{tab:#1}
                       \vspace{.1in}}

\newcommand {\btabular}[1]{\begin{center}
                       \begin{tabular}{#1}}
\newcommand {\etabular}{\end{tabular}
                       \end{center}}

\newcommand {\bdefin}[1]{\begin{definition}
                      \mymarginpar{def:#1}
                      \label{def:#1} }
\newcommand {\edefin}       {\end{definition}}
\newcommand {\rdef}[1]{Definition \ref{def:#1}}

\newcommand {\bpro}[1]{\begin{property}
                      \mymarginpar{pro:#1}
                      \label{pro:#1} }
\newcommand {\epro}   {\end{property}}

\newcommand {\bprop}[1]{\begin{proposition}
                      \mymarginpar{prop:#1}
                      \label{prop:#1} }
\newcommand {\eprop}       {\end{proposition}}
\newcommand {\rprop}[1]{Proposition \ref{prop:#1}}

\newcommand {\blem}[1]{\begin{lemma}
                      \mymarginpar{lem:#1}
                      \label{lem:#1} }
\newcommand {\elem}   {\end{lemma}}
\newcommand {\rlem}[1]{Lemma \ref{lem:#1}}

\newcommand {\bthe}[1]{\begin{theorem}
                      \mymarginpar{the:#1}
                      \label{the:#1} }
\newcommand {\ethe}   {\end{theorem}}
\newcommand {\rthe}[1]{Theorem \ref{the:#1}}

\newcommand {\bproof}{\noindent {\bf Proof.} \ }
\newcommand {\eproof} {\hfill \squares \\ \vspace{.3cm}}
\newcommand {\bcor}[1]{\begin{corollary}
                      \mymarginpar{cor:#1}
                      \label{cor:#1} }
\newcommand {\ecor}   {\end{corollary}}
\newcommand {\rcor}[1]{Corollary \ref{cor:#1}}

\newcommand {\bax}[1]{\begin{axiom}
                      \mymarginpar{ax:#1}
                      \label{ax:#1} }
\newcommand {\eax}       {\vspace{-.1in} \end{axiom}}

\newcommand {\bex}[2]{\vspace{.1in}
                      \begin{example}
                      \mymarginpar{ex:#1}
                       {\bf #2}
                      \label{ex:#1} }
\newcommand {\eex}       {\end{example} \vspace{.3cm} }
\newcommand {\rex}[1]{Example \ref{ex:#1}}

\newcommand {\brem}[1]{\begin{remark}
                      \mymarginpar{rem:#1}
                      \label{rem:#1} \em }
\newcommand {\erem}   {\end{remark}}

\newcommand {\beq}[1]{\mymarginpar{eq:#1}
                      \begin{equation}
                      \label{eq:#1} }

\newcommand {\beqno}[1]{\mymarginpar{eq:#1}
                      \begin{eqnarray}
                      \nonumber}

\newcommand {\eeq}       {\end{equation}}
\newcommand {\eeqno}       { && \end{eqnarray}}
\newcommand {\req}[1]{(\ref{eq:#1})}

\newcommand {\bear}[1]{\mymarginpar{eq:#1}
                       \begin{eqnarray}
                       \label{eq:#1} }

\newcommand {\bearno}[1]{\mymarginpar{eq:#1}
                       \begin{eqnarray}
                       \nonumber}

\newcommand {\eear}{\end{eqnarray}}
\newcommand {\eearno}{\end{eqnarray}}
\newcommand {\bsel}{\left \{ \begin{array}{cl}}
\newcommand {\esel}{\end{array} \right.}

\newcommand {\bmat}[1]{\left [ \begin{array}{#1}}
\newcommand {\emat}{\end{array} \right ]}
\newcommand {\bsec}[2]{\mymarginpar{sec:#2}
                       \section{#1}
                       \label{sec:#2} }

\newcommand {\rsec}[1]{Section \ref{sec:#1}}

\newcommand {\bsubsec}[2]{\mymarginpar{sec:#2}
                       \subsection{#1}
                       \label{sec:#2} }

\def\R{I\kern-0.30em R}
\def\N{I\kern-0.30em N}
\def\P{I\kern-0.30em P}
\newcommand\squares{\vrule height6pt width7pt depth1pt}

\def\pr{{\bf\sf P}}

\newcommand{\rhog}{\rho}
\newcommand{\rone}{\tau}

\newcommand{\qinf}{q^{(\infty)}_{\bf 0}}
\newcommand{\trhoj}{\rho^{(j)}}
\newcommand{\Rj}{R^{(j)}}

\begin{document}

\title{On the Stability Regions of Coded Poisson Receivers with Multiple Classes of Users and Receivers}

%
\author{Chia-Ming Chang, Yi-Jheng Lin, Cheng-Shang~Chang,~\IEEEmembership{Fellow,~IEEE,}\\ and Duan-Shin Lee,~\IEEEmembership{Senior Member,~IEEE}
                \thanks{C.-M. Chang, Y.-J. Lin, C.-S. Chang, and D.-S. Lee are with the Institute of Communications Engineering, National Tsing Hua University, Hsinchu 30013, Taiwan, R.O.C. Email:  jamie@gapp.nthu.edu.tw;  s107064901@m107.nthu.edu.tw;   cschang@ee.nthu.edu.tw;  lds@cs.nthu.edu.tw.  This work was supported in part by the Ministry of
Science and Technology, Taiwan, under Grant 109-2221-E-007-091-MY2, and in part by Qualcomm Technologies under Grant SOW NAT-435533.}
}

\maketitle

\begin{abstract}
Motivated by the need to provide differentiated quality-of-service (QoS) in grant-free uplink transmissions in 5G networks and beyond, we extend the probabilistic analysis of coded Poisson receivers (CPR) to the setting with multiple classes of users and receivers.
For such a CPR system, we prove (under certain technical conditions) that there is a region, called the {\em stability region} in this paper.  Each transmitted packet can be successfully received with probability 1 when the offered load to the system is within the stability region. On the other hand, if the offered load is outside the stability region, there is a nonzero probability that a packet will fail  to be received. We then extend the stability region to the $\epsilon$-stability region for CPR systems with decoding errors. We also demonstrate the capability of providing  differentiated QoS in such CPR systems by comparing the stability regions under various parameter settings.
\end{abstract}


\section{Introduction}
\label{sec:introduction}

One of the most challenging problems for the fifth-generation networks (5G) and beyond is to
support various connectivity classes of users, including
(i) enhanced mobile broadband (eMBB), (ii) ultra-reliable low-latency communications (URLLC), and (iii) massive machine-type communications (mMTC) (see, e.g., \cite{li20175g,bennis2018ultra,popovski2019wireless,le2020overview} and references therein). These connectivity classes have different requirements. For instance, to provide the URLLC services for
autonomous driving, drones, and augmented/virtual reality, the  reliability defined in 3GPP requires the $1-10^{-5}$ success probability of transmitting a layer two protocol data unit of 32 bytes within 1 ms. For some use cases, such as electrical power distribution, might have stricter requirements (higher reliability of $10^{-6}$, lower latency of
0.5 to 1 ms) \cite{le2020overview}.
For downlink transmissions,
these quality-of-service (QoS) requirements could be supported by centralized scheduling algorithms. In particular,  utility-based algorithms  were proposed in \cite{anand2020joint} for joint scheduling of URLLC and
  eMBB traffic. However, it is much more difficult to meet various QoS requirements for grant-free uplink transmissions as this has to be done in a distributed manner for a multiple access channel (that depends heavily on the physical layer).

There have been many multiple access schemes proposed in the literature, see, e.g.,
\cite{casini2007contention,liva2011graph,narayanan2012iterative,paolini2012random,jakovetic2015cooperative,sun2017coded,stefanovic2018coded,Hoshyar2008,SCMA,Yuan2016,Chen2017,ordentlich2017low,vem2019user,andreev2020polar}.
Most of these schemes are heavily correlated to various physical layer characteristics, including encoding/decoding schemes, fading channel models, modulation schemes, and power allocations. As such, it is very difficult to extend the analyses in these multiple access schemes to a higher layer protocol, where there are multiple classes of users and receivers.

Recently, an abstract receiver at the Medium Access Control (MAC) layer, called {\em Poisson receiver}, was proposed in \cite{chang2020Poisson} to address the QoS requirements for various grant-free uplink transmissions. A Poisson receiver specifies the probability that a packet is successfully received (decoded) when the number of packets  transmitted simultaneously to the receiver follows a Poisson distribution (Poisson offered load).
As such, one can hide the encoding/decoding complexity from the physical layer by summarizing the physical layer as an input-output function of the (packet) success probability. Like Irregular Repetition Slotted
ALOHA (IRSA) \cite{liva2011graph} and coded slotted ALOHA (CSA) \cite{narayanan2012iterative,paolini2012random,jakovetic2015cooperative,sun2017coded,stefanovic2018coded}, one can also group $T$ independent Poisson receivers together to form a system of coded Poisson receivers (CPR). In a CPR system,
a random number of copies of each packet (from an active  user) are transmitted uniformly to the $T$ receivers. If any one of these  copies of a packet is successfully received by a receiver, then the other copies can be removed from the system to further improve the success probabilities of the other packets. Such a process can  then be iteratively  carried out for $i$ times to ``decode'' the rest of the packets.
 As $T$ and $i$ go to infinity, several numerical examples in \cite{chang2020Poisson} suggest that such an iterative decoding approach leads to a Poisson receiver with a success probability function of threshold type. When the offered load is below the threshold, each packet can be successfully received with probability 1. On the other hand, when the offered load exceeds the threshold, there is a nonzero probability that a packet will fail to be received. Such a percolation phenomenon was reported earlier in IRSA \cite{liva2011graph} and CSA \cite{narayanan2012iterative,paolini2012random,jakovetic2015cooperative,stefanovic2018coded}.

To provide differentiated QoS for multiple classes of users in 5G networks, one common approach, known as {\em network slicing} \cite{popovski20185g}, is to partition and allocate available radio resources  in an efficient manner to users. On the other hand, to meet the stringent QoS guarantees for URLLC uplink transmissions,
 the Third Generation Partnership Project (3GPP) defines the {\em configured-grant transmissions} that allow URLLC traffic to transmit multiple copies of a packet in a frame \cite{le2020overview,le2021enhancing}.
 Motivated by these,
 we extend the analysis for the CPR system in \cite{chang2020Poisson} to the setting with multiple classes of users and receivers.
 Specifically, we partition the $T$ receivers (radio resources) into $J$ classes. The class $j^{th}$ receivers are allocated with $F_j T$ receivers, where $\sum_{j=1}^J F_j=1$.
Unlike the CPR system with a single class of receivers in \cite{chang2020Poisson}, copies of packets are no longer transmitted {\em uniformly} to the $T$ receivers. Suppose that there are $K$ classes of users in the CPR system. Then
each copy of a  packet from a  class $k$  user is transmitted  to  class $j$ receivers  with the routing probability $r_{k,j}$. The partition of the $T$ receivers and the routing probabilities specify how the $T$ receivers are shared among the users.
Under certain technical assumptions, we prove that for such a CPR system with multiple classes of users and receivers,
there exists a region, called the {\em stability region} in this paper,  such that each packet can be successfully received with probability 1 when the offered load is within the region. On the other hand, if the offered load is outside the stability region, then there is a nonzero probability that a packet will fail to be received.
The stability region provides a natural way for admission control of configured-grant transmissions.

The stability region that ensures every packet is successfully received may not hold in a practical setting, where
there are noise, channel fading, and decoding errors. To take these practical factors into account, we further extend the stability region to the $\epsilon$-stability region. The vector $\epsilon=(\epsilon_1,\epsilon_2,\ldots, \epsilon_K)$ is a vector of $K$ parameters that are chosen to meet the QoS requirements of the $K$ classes of users. When the offered load is within the $\epsilon$-stability region, a packet, depending on its class, can be successfully received with a guaranteed probability in terms of $\epsilon$.

By specifying the routing probabilities, one has the flexibility to either enlarge or shrink the stability region (and the $\epsilon$-stability region) for certain classes of users.
To demonstrate this, we consider an IRSA system with two classes of users and two classes of receivers and compare the stability regions under four packet routing policies: (i) complete sharing (where the receivers are shared equally),
(ii) receiver reservation (where one class of receivers are reserved for one class of users), (iii) (nearly) complete partitioning (where receivers are not shared), and (iv) nonuniform sharing (where receivers are shared unequally).
The complete sharing policy has a convex stability region. However, the stability region of the receiver reservation policy is not convex. Moreover, for the class of users that have reserved  access for one class of receivers, the receiver reservation policy can accommodate a larger offered load than that of the complete sharing policy.
 As the receivers are not shared under the (nearly) complete partitioning policy, its performance is the worst among these four policies, and it has the smallest stability region. When there are decoding errors in the IRSA system, the notion of stability region is no longer valid. Instead,  we investigate the $\epsilon$-stability region and show the trade-off between the admissible offered load to the system and the QoS requirement for the packet success probability.

 In addition to the IRSA system, we also consider the CPR system with each Poisson receiver modelled by the Rayleigh  block fading channel with capture in \cite{stefanovic2014exploiting,clazzer2017irregular,stefanovic2018coded}. Our numerical results show that the packet success probability (as a function of the offered load) is of threshold type.
 When the signal-to-noise ratio is very large, one can admit the load until the threshold without sacrificing  the QoS requirement for the packet success probability. However, when the signal-to-noise ratio is small,
 there is a clear trade-off between the admissible offered load  and the required packet success probability.


We summarize our contributions as follows:

\noindent (i) We extend the analysis of CPR systems in \cite{chang2020Poisson} to the setting with multiple classes of users and receivers. This is done by introducing a partition of receivers and routing probabilities from various classes of users to various classes of receivers into  CPR systems. Our work can also be extended to
the independent packet erasure channel in \cite{sun2017coded}, and the independent SIC error model in \cite{liva2011graph,dumas2021design}.

\noindent (ii) We define the notion of stability for such CPR systems and prove the existence of stability regions.
We also derive sufficient conditions to characterize the stability regions.

\noindent (iii) We extend the notion of stability to the notion of $\epsilon$-stability to cope with the practical setting with noise, channel fading, and decoding errors.

\noindent (iv) By conducting extensive numerical studies, we show how the stability regions and the $\epsilon$-stability regions are affected by various parameters of CPR systems, including routing probabilities and degree distributions. Fine tuning these parameters can lead to differentiated QoS among multiple classes of users.

In this paper, we
let ${\bf 0}$ (resp. ${\bf 1}$) be the $1 \times K$ vector with all its elements being 0 (resp. 1).
Also, for two vectors $x$ and $y$, we say $x \le y$ if every element in $x$ is not larger than the corresponding element in $y$. A list of notations is given in Appendix A of the supplementary material.

The rest of the paper is organized as follows.
In \rsec{poisson}, we briefly review the framework of Poisson receivers in  \cite{chang2020Poisson}. We then extend the framework to the setting with multiple classes of users and receivers in \rsec{mulr}. We define the notion of stability and develop its associated theoretical analyses  in \rsec{stability}. In \rsec{num}, we provide numerical results for the stability regions and the throughputs of various CPR systems. The paper is concluded in \rsec{con}.

\section{Review of the framework of Poisson receivers}
\label{sec:poisson}


In this section, we briefly review the framework of Poisson receivers in  \cite{chang2020Poisson}.
We say a system with $K$ classes of input traffic is subject to a Poisson offered load  $\rho=(\rho_1, \rho_2, \ldots, \rho_K)$ if these $K$ classes of input traffic are {\em independent}, and the number of class $k$ packets arriving at the system
follows a Poisson distribution with mean $\rho_k$, for $k=1,2, \ldots, K$.


\bdefin{Poissonmul}{(\bf Poisson receiver with multiple classes of input traffic \cite{chang2020Poisson})}
An abstract receiver
is called a {\em $(P_{{\rm suc},1}(\rhog), P_{{\rm suc},2}(\rhog), \ldots, P_{{\rm suc},K}(\rhog))$-Poisson receiver} with $K$ classes of input traffic if the receiver is subject to a Poisson offered load  $\rho=(\rho_1, \rho_2, \ldots, \rho_K)$, a tagged (randomly selected) class $k$  packet
is successfully received with probability $P_{{\rm suc},k}(\rhog)$, for $k=1,2, \ldots, K$.
\edefin

The throughput of class $k$ packets  (defined  as the expected number of class $k$ packets that are successfully received) for a {\em $(P_{{\rm suc},1}(\rhog), P_{{\rm suc},2}(\rhog), \ldots, P_{{\rm suc},K}(\rhog))$-Poisson receiver} subject to a Poisson offered load $\rho$ is thus
\beq{Poithrmul}
\Theta_k=\rho_k \cdot P_{{\rm suc},k}(\rhog),
\eeq
$k=1,2, \ldots, K$.

It was shown in \cite{chang2020Poisson} that many systems could be modelled by Poisson receivers, including Slotted ALOHA (SA) \cite{ALOHA} and SA with multiple {\em cooperative} receivers.
Moreover, there are two elegant closure properties of Poisson receivers  for packet routing and packet coding in \cite{chang2020Poisson}.
These two closure properties greatly reduce the computational complexity of the density evolution method
\cite{gallager1962low,luby1998analysisb,richardson2001capacity} used for analyzing CSA, and they
 serve as building blocks for analyzing a large system of Poisson receivers.

In the following, we give four examples of Poisson receivers in \cite{chang2020Poisson,liu2020aloha} that will be used for evaluating our numerical results.

\bex{TfoldALOHA}{(D-fold ALOHA)}
$D$-fold ALOHA  \cite{ordentlich2017low,stefanovic2017asymptotic,glebov2019achievability,vem2019user} is a generalization of the SA system, where the maximum number of packets that can be successfully received is $D$.
It was shown in  \cite{chang2020Poisson} that $D$-fold ALOHA is a Poisson receiver with
\beq{tfold1111}
P_{\rm suc}(\rhog)=\sum_{t=0}^{D-1} \frac{e^{-\rho} \rho^{t}}{t!}.
\eeq
\eex

\bex{errorALOHA}{($D$-fold ALOHA with decoding errors)}
In practice, there are decoding errors.  Suppose that the probability of decoding errors is $p_{\rm err}$
in a $D$-fold ALOHA. Then we can model it by a mixture of two Poisson receivers. With probability $1-p_{\rm err}$, it is a Poisson receiver with the success probability function in \req{tfold1111}. On the other hand,  with probability $p_{\rm err}$, it is a Poisson receiver with the success probability function equal to 0.
It is easy to see such a mixture of two Poisson receivers is a Poisson receiver with
the following success probability function:
\beq{tfold1111mix}
P_{\rm suc}(\rhog)
=(1-p_{\rm err}) \sum_{t=0}^{D-1} \frac{e^{-\rho} \rho^{t}}{t!}.
\eeq
Such a model for  $D$-fold ALOHA with decoding errors was previous shown in \cite{liu2020aloha}. For $D=1$, it is known as the slot erasure model in \cite{sun2017coded}.
\eex

\bex{tworeceiversb}{(Two cooperative SA receivers)}
In \cite{chang2020Poisson,liu2020aloha}, they
considered a system
with two cooperative SA receivers that can exchange the information of successfully received packets from each other to perform successive inference cancellation (SIC). There are two classes of traffic: the URLLC traffic (class 3 in \cite{chang2020Poisson,liu2020aloha}) and the eMBB traffic (class 4 in \cite{chang2020Poisson,liu2020aloha}). Each packet from the URLLC traffic is sent to both receivers. On the other hand, a packet from the eMBB traffic is sent to  one of the two receivers with an equal probability.
Suppose that the URLLC traffic has a Poisson offered load $\rho_3$ and the eMBB traffic has a Poisson offered load $\rho_4$.
Then such a system is a Poisson receiver with the success probability functions:
\beq{coop1111}
P_{{\rm suc},3}(\rho_3,\rhog_4)=2e^{-(\rho_3+\frac{1}{2}\rho_4)}-e^{-(\rho_3+\rho_4)},
\eeq
and
\beq{coop1111a}
P_{{\rm suc},4}(\rho_3,\rhog_4)=e^{-(\rho_3+\frac{1}{2}\rho_4)}+\rho_3e^{-(\rho_3+\rho_4)}.
\eeq
\eex

\bex{Rayleigh}{(Rayleigh  block fading channel with capture)}
In \cite{liu2020aloha}, it was shown that
the Rayleigh block fading channel with capture in \cite{stefanovic2014exploiting,clazzer2017irregular,stefanovic2018coded}
could be modelled by a Poisson receiver.
In a wireless channel with $N$ active users and one receiver,
the packet of the $k^{th}$ user can be successfully received if
the signal-to-interference-and-noise ratio (SINR) is
higher than a predefined threshold.
In particular, for the Rayleigh block fading channel, such a condition can be written as
\beq{fading3333r}
\frac{X_k}{\sum_{n \ne k} X_n +\frac{1}{\gamma}} \ge b,
\eeq
where $X_n$'s are independent and exponentially distributed with mean 1, $\gamma$ is the signal-to-noise ratio, and $b$ is a predefined threshold.
As such, the probability that the packet of the $k^{th}$ user can be successfully received  (see, e.g., \cite{stefanovic2014exploiting,clazzer2017irregular,stefanovic2018coded}) is
\beq{fading4444r}
\pr (\frac{X_k }{\sum_{n \ne k} X_n  +\frac{1}{\gamma}} \ge b)=\frac{e^{-b/\gamma}}{(1+b)^{N-1}}.
\eeq

For the capture effect in the threshold-based model,
we can use the SIC decoding algorithm. As shown in \cite{liu2020aloha},
the average number of packets that can be successfully received when there are $N$ active users
is
\bear{fading6666r}
\sum_{r=1}^N \frac{N!}{(N-r)!}\frac{e^{-\frac{1}{\gamma}((1+b)^{r}-1)}}{(1+b)^{r\left(N-1-\frac{r-1}{2}\right)}}.
\eear
This is then used to show that the Rayleigh  block fading channel with capture is a $P_{\rm suc}(\rho)$-Poisson receiver with
\bear{fading8888r}
P_{\rm suc}(\rho)
&=&\sum_{t=0}^\infty  \sum_{\rone=0}^{t} \frac{e^{-\rho} \rho^t}{(t-\rone)!} \frac{e^{-\frac{1}{\gamma}((1+b)^{\rone+1}-1)}}{(1+b)^{(\rone+1)\left(t-\frac{\rone}{2}\right)}}.
\eear
\eex

\bsec{Coded Poisson receivers with multiple classes of users and receivers}{mulr}

\begin{figure}[ht]
	\centering
	\includegraphics[width=0.43\textwidth]{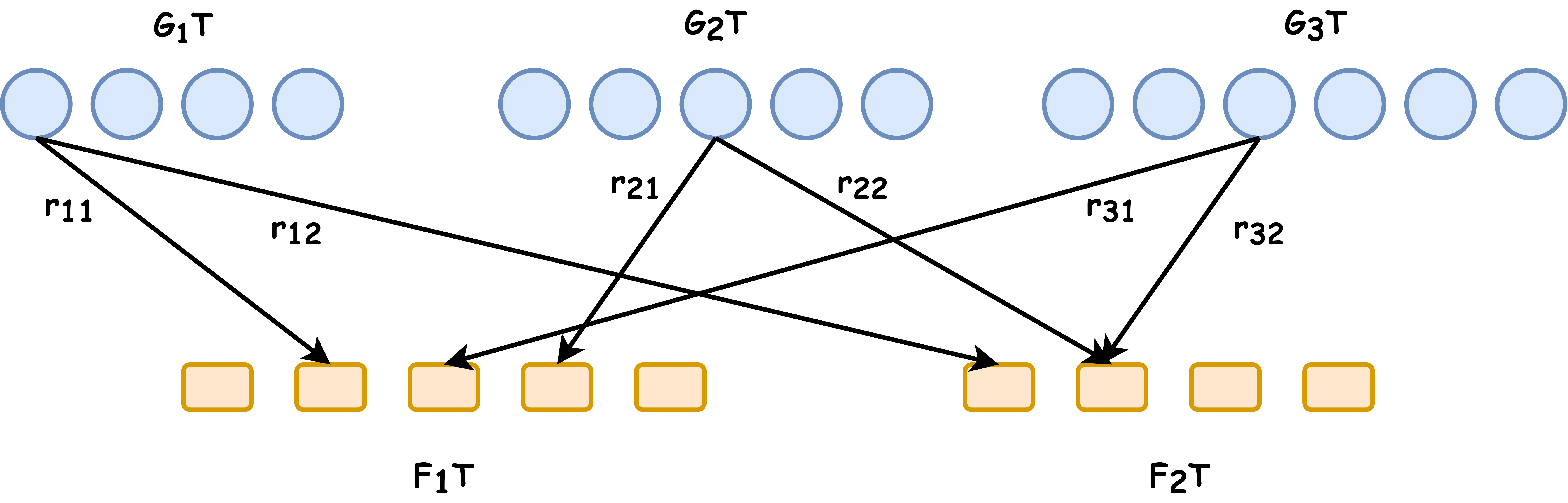}
	\caption{A bipartite graph representation of SIC decoding for three classes of users and two classes of Poisson receivers.}
	\label{fig:pdecoding}
\end{figure}

The main objective of this section is to extend the result for coded Poisson receivers with a {\em single} class of receivers in \cite{chang2020Poisson} to the setting with {\em multiple} classes of  receivers.
Specifically, we consider a system with $G_{k} T$ class $k$ active users, $k=1,2, \ldots, K$, and $F_j T$ class $j$
Poisson receivers, $j=1,2, \ldots, J$.  Class $j$ Poisson receivers have the success probability functions $P_{{\rm suc},1,j}(\rhog), P_{{\rm suc},2,j}(\rhog), \ldots, P_{{\rm suc},K,j}(\rhog)$ for the $K$ classes of input traffic. Each class $k$ user transmits its packet  for $L_k \ge 1$ times (copies). With the routing probability $r_{k,j}$ ($\sum_{j=1}^J r_{k,j}=1$), each copy of a class $k$ packet is transmitted  {\em uniformly} and {\em independently}
  to one of the $F_j T$ class $j$ Poisson receivers.
For such a system, we assume perfect SIC, i.e., as long as one copy of a packet is successfully received by one of the receivers, then it can be used to remove the other copies of that packet from the other receivers. Moreover,
all the Poisson receivers are assumed to be {\em independent}, i.e., the event that a packet is successfully received by a Poisson receiver  is independent of the outcomes of the other Poisson receivers as long as their input traffic is independent of each other.
As  in \cite{chang2020Poisson}, the decoding process can be described by a peeling decoder on a user-receiver bipartite graph, where an edge between a user node and a receiver node is added if there is a packet transmitted from that user node to that receiver node.
Call an edge a class $k$ edge if the user end of the edge is connected to a class $k$ user. Also, we
call an edge a class $(k,j)$-edge if the receiver (resp. user) end of the edge is connected to a class $j$ receiver (resp. class $k$ user).
To illustrate this, we show in \rfig{pdecoding} a system of three classes of users and two classes of Poisson receivers.


As in \cite{chang2020Poisson}, we let $\Lambda_{k,\ell}$ be the probability that a class $k$ packet is transmitted $\ell$ times, i.e.,
\beq{extm1111}
P(L_{k}=\ell)=\Lambda_{k,\ell}, \;\ell=1,2,\dots
\eeq
Define the generating function
\beq{mean0000mulr}
\Lambda_{k}(x)=\sum_{\ell=0}^\infty \Lambda_{k,\ell} \cdot x^\ell
\eeq
 of the {\em degree distribution} of a class $k$ user node, and
 the generating function
 \beq{mean3333mulr}
\lambda_{k}(x)=\sum_{\ell=0}^\infty \lambda_{k,\ell} \cdot x^\ell
\eeq
of the {\em excess degree distribution} of a class $k$ user node, where
\beq{mean2222mulr}
\lambda_{k,\ell}=\frac{\Lambda_{k,\ell+1}\cdot (\ell+1)}{\sum_{\ell=0}^\infty\Lambda_{k,\ell+1}\cdot (\ell+1)}
\eeq
is the probability that the  user end of a randomly selected class $k$ edge has additional $\ell$ edges excluding the randomly selected class $k$  edge.
Note that the mean degree of a class $k$ user node is
\beq{mean1111mulr}
\Lambda_{k}^{\prime}(1)=\sum_{\ell=0}^\infty \ell\cdot \Lambda_{k,\ell},
\eeq
and that
\beq{mean3344mulr}
\lambda_{k}(x)=\frac{\Lambda_{k}^\prime(x)}{\Lambda_{k}^{\prime}(1)}.
\eeq

Our analysis is based on the tree evaluation method in \cite{luby1998analysis,liva2011graph,paolini2011graph,paolini2012random} and the reduced Poisson offered load argument in \cite{chang2020Poisson}. The tree evaluation method
was extended in \cite{luby1998analysisb,richardson2001capacity} to derive the capacity of low-density parity-check (LDPC) codes \cite{gallager1962low}, and it is commonly known as the density evolution (DE) method to the information theory community. In fact, it was pointed out in \cite{i2018finite} a very interesting connection between IRSA and high-rate LDPC codes.
On the other hand, the reduced Poisson offered load argument is a widely used method to analyze queueing networks (see, e.g., \cite{kelly2011reversibility,walrand1983probabilistic,kelly1991loss}).

It consists of the following steps:

\noindent (i)
The initial offered load of class $k$ packets to a class $j$ Poisson receiver, defined as the expected number of class $k$ packets transmitted to that receiver, is
\beq{mean4444mulr}
\rho_{k,j}= G_{k} \Lambda_{k}^\prime(1) r_{k,j}/F_j.
\eeq
To see \req{mean4444mulr}, note that (a) there are $G_k T$ class $k$ users, (b) each class $k$ user transmits on average $\Lambda_{k}^\prime(1)$ copies, (c) each copy is sent to  class $j$ Poisson receivers with the routing probability $r_{k,j}$, and (d) a copy sent to class $j$ Poisson receivers is uniformly distributed among the $F_j T$ class $j$ Poisson receivers.
When $T$ goes to infinity, the number of class $k$ packets at a  class $j$ receiver converges (from a binomial random variable) to
a Poisson random variable with mean $\rho_{k,j}$, and the degree distribution of class $k$ packets at a class $j$ receiver node is a Poisson distribution with  mean $\rho_{k,j}$.

\noindent (ii) Let
$q_{k}^{(i)}$ be
the probability that  the  {\em user end} of a randomly selected class $k$ edge has not been successfully received after the $i^{th}$ SIC iteration. The
offered load of class $k$ packets to a class $j$ Poisson receiver after the $i^{th}$ SIC iteration
has a Poisson distribution with  mean $q_{k}^{(i)} \rho_{k,j}$. As pointed out in \cite{chang2020Poisson}, this is due to two important  closure properties of Poisson random variables: (a) randomly splitting a Poisson random variable yields independent Poisson random variables, and (b) the excess degree distribution of a Poisson random variable is Poisson with the same mean.
As such, the Poisson offered load is  reduced from $\rho_{k,j}$ to $q_{k}^{(i)} \rho_{k,j}$ after the $i^{th}$ SIC iteration.
Let
\beq{rho0000r}
\trhoj=(\rho_{1,j}, \rho_{2,j}, \ldots, \rho_{K,j}),
\eeq
and
\beq{rhoiiiir}
q^{(i)}=(q_{1}^{(i)}, q_{2}^{(i)}, \ldots, q_{K}^{(i)}).
\eeq
We can represent  the offered load at a class $j$ Poisson receiver after the $i^{th}$ SIC iteration
by the vector $q^{(i)} \circ \trhoj$, where
$\circ$ denotes the element-wise multiplication of two vectors. This reduced load argument is also the key step that greatly reduces the computational complexity in the density evolution method.

\noindent (iii) Let $p_{k,j}^{(i+1)}$ be the probability that  the {\em receiver  end} of a randomly selected class $(k,j)$-edge has not been successfully received after the $(i+1)^{th}$ SIC iteration. Then
\beq{tag6666amulr}
p_{k,j}^{(i+1)}=1- P_{{\rm suc},k,j}(q^{(i)} \circ \trhoj).
\eeq
That \req{tag6666amulr} holds follows directly from the definition of a Poisson receiver in \rdef{Poissonmul} as
the offered load at a class $j$ Poisson receiver after the $i^{th}$ SIC iteration
is $q^{(i)} \circ \trhoj$.

\noindent (iv) Let $p_{k}^{(i+1)}$  be
the probability that  the {\em receiver end} of a randomly selected class $k$ edge has not been successfully received after the $(i+1)^{th}$ SIC iteration. Since a class $k$ edge is a class $(k,j)$-edge with probability $r_{k,j}$, it follows that
\bear{tag6666bmulr}
p_k^{(i+1)}&=&\sum_{j=1}^J r_{k,j}p_{k,j}^{(i+1)} \nonumber\\
&=&1-\sum_{j=1}^J r_{k,j}P_{{\rm suc},k,j}(q^{(i)} \circ\trhoj).
\eear

\noindent (v) The probability $q_k^{(i)}$ can be computed recursively from the following equation:
\beq{tag6666cmulr}
q_{k}^{(i+1)}=\lambda_{k}(1- \sum_{j=1}^J r_{k,j}P_{{\rm suc},k,j}(q^{(i)} \circ \trhoj)),
\eeq
with $q_{k}^{(0)}=1$.
To see this,
note that a packet sent from a user (the user end of the bipartite graph) can be successfully received if at least one of its copies is successfully received at the {\em receiver} end.
Since the probability that the user end of a randomly selected class $k$ edge  has additional $\ell$ edges is $\lambda_{k,\ell}$, the probability that the {\em user} end of a randomly selected class $k$ edge  cannot be successfully received after the $(i+1)^{th}$ iteration is
\bear{tag2222mulr}
q_{k}^{(i+1)}&=&1-\sum_{\ell=0}^\infty \lambda_{k,\ell} \cdot \Big (1-(p_{k}^{(i+1)})^{\ell} \Big) \nonumber\\
&=&\lambda_{k}(p_{k}^{(i+1)}).
\eear
Using \req{tag6666bmulr} in \req{tag2222mulr} yields \req{tag6666cmulr}.

\noindent (vi) Let $\tilde P_{{\rm suc},k}^{(i)}$ be
the probability that a packet sent from a randomly selected {\em class $k$ user} can be successfully received after the $i^{th}$ iteration.
   Such a probability is the probability that at least one copy of the packet has been successfully received after the $i^{th}$ iteration. Since the probability that a randomly selected {\em class $k$ user} has  $\ell$ edges is $\Lambda_{k,\ell}$, we have from \req{tag6666bmulr} that
\bear{mean5555dmulr}
&&\tilde P_{{\rm suc},k}^{(i)}=\sum_{\ell=0}^\infty \Lambda_{k,\ell} \cdot \Big (1-(p_{k}^{(i)})^{\ell} \Big)\nonumber\\
 &&=\sum_{\ell=0}^\infty \Lambda_{k,\ell} \cdot \Big (1-(1- \sum_{j=1}^J r_{k,j}P_{{\rm suc},k,j}(q^{(i-1)} \circ \trhoj))^{\ell} \Big) \nonumber \\
&&=1-\Lambda_k\Big (1- \sum_{j=1}^J r_{k,j}P_{{\rm suc},k,j}(q^{(i-1)} \circ \trhoj)\Big).
\eear

Let $$G=(G_1, G_2, \ldots, G_{K}),$$
$$\Lambda^\prime (x)=(\Lambda^\prime_1 (x), \Lambda^\prime_2 (x), \ldots, \Lambda^\prime_K (x)),$$
and
\beq{mulm4444}
\Rj=(\frac{r_{1,j}}{F_j}, \frac{r_{2,j}}{F_j}, \ldots, \frac{r_{K,j}}{F_j})
\eeq
for $j=1,2, \ldots, J$.
In view of \req{mean4444mulr} and \req{rho0000r}, we have
\beq{mean6666dmulr}
\trhoj= G \circ \Lambda^\prime(1)\circ \Rj.
\eeq

We summarize the result in \req{mean5555dmulr} in the following theorem.

\bthe{mainext}
As $T \to \infty$,
the system of coded Poisson receivers after the $i^{th}$ SIC iteration converges to a Poisson receiver with the success probability function for the class $k$ traffic
\bear{mean8888thumulam}
 &&\tilde P_{{\rm suc},k}^{(i)}(G)
 =1-\Lambda_k \Big  (1- \nonumber\\
 &&\sum_{j=1}^J r_{k,j} P_{{\rm suc},k,j}(q^{(i-1)} \circ G \circ \Lambda^\prime(1) \circ \Rj)\Big),
\eear
$k=1,2, \ldots, K$, where $q^{(i)}=(q_{1}^{(i)}, q_{2}^{(i)}, \ldots, q_{K}^{(i)})$ can be computed recursively from the following equation:
\beq{tag6666dmulm}
q_{k}^{(i+1)}=\lambda_{k}\Big (1-\sum_{j=1}^J r_{k,j} P_{{\rm suc},k,j}(q^{(i)} \circ G \circ \Lambda^\prime(1) \circ \Rj)\Big ),
\eeq
with $q^{(0)}=(1, 1, \ldots, 1)$.
\ethe

We note that here we only outline the six steps to derive the results in \rthe{mainext}.
A rigorous proof for  \rthe{mainext}  requires the concentration theorem (Theorem 2 of \cite{richardson2001capacity}) to prove the ergodicity of the system, i.e., as $T \to \infty$,
the average fraction of nodes that have not been successfully received after the $i^{th}$ SIC iteration converges to the probability that a randomly selected node has not been successfully received after the $i^{th}$ SIC iteration.

In comparison with the density evolution method in \cite{Liva2012spatially,sandgren2016frame} for analyzing spatially-coupled random access, the explicit functional representation in \rthe{mainext} is much more succinct. We will use the functional representation to analyze the stability of CPR systems via the solutions of a set of functions in the next section.

There are two simple extensions for the results in \rthe{mainext}: (i) independent packet erasure channel \cite{sun2017coded}, and (ii) independent SIC error model \cite{liva2011graph,dumas2021design}. For the independent packet erasure channel, each packet is erased independently (due to fading) with probability $p_{\rm era}$. For such a model, the number of copies is simply a random sum of independent Bernoulli random variables, and
the new generating function of the degree distribution of class $k$ packets is
\beq{erase1111}\tilde \Lambda_k(x)= \Lambda_k(p_{\rm era}+(1-p_{\rm era})x),
\eeq
where $\Lambda_k(x)$ is the original generating function of the degree distribution of class $k$ packets.
This was previous shown in Eq. (13) of \cite{sun2017coded}.

For the independent SIC error model, a copy of a successfully received packet cannot be removed from the SIC operation with probability  $p_{\rm sic}$. Note that in the perfect SIC setting, we simply have $p_{\rm sic}=0$. In such a model, the mean
offered load of class $k$ packets to a class $j$ Poisson receiver after the $i^{th}$ SIC iteration
is no longer $q_{k}^{(i)} \rho_{k,j}$. By taking the SIC error into account, this is increased to
$(p_{\rm sic}+(1-p_{\rm sic})q_{k}^{(i)}) \rho_{k,j}$. Thus, one only needs to replace $q^{(i)}$
in \req{tag6666dmulm} by
$ p_{\rm sic} {\bf 1}+ (1-p_{\rm sic})q^{(i)}$.
Also, $q^{(i-1)}$ in
\req{mean8888thumulam} needs to be changed accordingly. Note that the probability that the $(\ell-1)^{th}$ SIC operation is successful in the independent SIC error model is $(1-p_{\rm sic})^{\ell-1}$, which is exactly the same as Eq. (1) in \cite{dumas2021design}.

\bsec{Stability}{stability}

In this section, we discuss the stability region of coded Poisson receivers with multiple classes of input traffic.

\bdefin{srates}{(\bf Stability of coded Poisson receivers with multiple classes of input traffic)}
Consider the coded Poisson receiver described in \rsec{mulr}. A Poisson offered load $G=(G_1, G_2, \ldots, G_K)$ to the coded Poisson receiver is said to be {\em stable} if the probability that a packet is successfully received approaches to 1 when the number of iterations goes to infinity, i.e.,
\beq{srates1111}
\lim_{i \to\infty} \tilde P_{{\rm suc},k}^{(i)}(G)=1, \quad k=1,2,\ldots, K,
\eeq
where $\tilde P_{{\rm suc},k}^{(i)}(G)$ is defined in \req{mean8888thumulam}.
\edefin

The definition for the stability of coded Poisson receivers is motivated by the notion of stability in queueing theory, where a queue is (rate) stable if the departure rate is the same as the arrival rate.
For our analysis of the stability region, we consider the following assumptions:

\begin{description}
\item[(A1)]
For all $k=1, \ldots, K$ and $j=1,2, \ldots, J$, the success  probability function $P_{{\rm suc},k,j}(\rho)$ is  a continuous and decreasing function of $\rho$, and $P_{{\rm suc},k,j}({\bf 0})=1$,
\item[(A2)] If  $\rho \ne {\bf 0}$, then $P_{{\rm suc},k,j}(\rho)<1$ for all $k=1, \ldots, K$, and $j=1,2, \ldots, J$.
\item[(A3)] $r_{k,j}>0$ for all $k=1, \ldots, K$, and $j=1,2, \ldots, J$.
\end{description}

The condition in (A1) is quite intuitive. In order for every packet to be successfully received, the success probability function needs to be increased (continuously) to 1 when the offer load $\rho$ is decreased to 0.

\bsubsec{A necessary and sufficient condition for stability}{condition}

In the following theorem, we show a necessary and sufficient condition for a  Poisson offered load $G$ to be stable.

\bthe{stable}
Under (A1), a Poisson offered load $G$  is stable if
$q={\bf 0}$ is the unique solution  in $[0,1]^K$ of the following  $K$ equations:
\beq{stable0000}
q_{k}=\lambda_{k}\Big (1- \sum_{j=1}^J r_{k,j} P_{{\rm suc},k,j}(q \circ G \circ \Lambda^\prime(1) \circ \Rj)\Big ),
\eeq
$k=1,2, \ldots, K$.
On the other hand, under (A1), (A2), and (A3), a positive  Poisson offered load $G$ (with $G_k > 0$ for all $k$) is stable only if
$q={\bf 0}$ is the unique solution  in $[0,1]^K$ of the $K$ equations in \req{stable0000}.
\ethe

The proof of \rthe{stable} requires the results in the following lemma
that shows the limiting vector of $q^{(i)}$ is the largest solution of the set of $K$ equations in \req{stable0000}.


\blem{limit}
Under (A1), there exist $0\le q_k^{(\infty)} \le 1$, $k=1,2, \ldots, K$ such that
\beq{stable4444}
\lim_{i \to\infty} q_k^{(i)}=q_k^{(\infty)},
\eeq
and $q^{(\infty)}=(q_1^{(\infty)},q_2^{(\infty)}, \ldots, q_K^{(\infty)})$ is a solution of the  $K$ fixed point equations in \req{stable0000}. Moreover, it is the largest solution among all the solutions in $[0,1]^K$, i.e.,
if $\hat q=(\hat q_1, \hat q_2, \ldots, \hat q_K) \in [0,1]^K$ is another solution, then
$\hat q_k \le q_k^{(\infty)}$, $k=1,2,\ldots, K$.
\elem

\bproof

We first show by induction that $q^{(i+1)} \le q^{(i)}$ for all $i$.
As $q_k^{(0)}=1$ for all $k$, clearly we have $q^{(1)} \le q^{(0)}$.
Assume that $q^{(i)} \le q^{(i-1)}$ as the induction hypothesis.
Note that the generating function $\lambda_k(x)$ is increasing in $x$ for $x \in [0,1]$.
It then follows from \req{tag6666dmulm}, the assumption that  $P_{{\rm suc},k,j}(\rho)$ is  decreasing in $\rho$, and the
induction hypothesis
that
\bear{stable3333}
&&q_{k}^{(i+1)}\nonumber\\
&&=\lambda_{k}\Big (1- \sum_{j=1}^J r_{k,j} P_{{\rm suc},k,j}(q^{(i)} \circ G \circ \Lambda^\prime(1) \circ \Rj)\Big )\nonumber\\
&&\le \lambda_{k}\Big (1- \sum_{j=1}^J r_{k,j} P_{{\rm suc},k,j}(q^{(i-1)}\circ G \circ \Lambda^\prime(1) \circ \Rj)\Big )\nonumber\\
&&=q_{k}^{(i)}.
\eear

As $q_k^{(0)}=1$ and $q_k^{(i+1)} \le q_k^{(i)}$ for all $i$, the decreasing sequence converges to $q_k^{(\infty)}$, i.e.,
\beq{stable4444b}
\lim_{i \to\infty} q_k^{(i)}=q_k^{(\infty)}.
\eeq
Moreover,
we have from \req{tag6666dmulm} and the continuity of the success probability function $P_{{\rm suc},k,j}(\rho)$ and the generating function $\lambda_k(x)$ that
\bear{stable5555b}
&&q_{k}^{(\infty)}=\lim_{i \to\infty} q_k^{(i+1)}\nonumber\\
&&=\lim_{i \to\infty}\lambda_{k}\Big (1- \sum_{j=1}^J r_{k,j} P_{{\rm suc},k,j}(q^{(i)} \circ G \circ \Lambda^\prime(1) \circ \Rj)\Big )\nonumber\\
&&=\lambda_{k}\Big (1- \sum_{j=1}^J r_{k,j} P_{{\rm suc},k,j}(q^{(\infty)} \circ G \circ \Lambda^\prime(1) \circ \Rj)\Big ).\nonumber\\
\eear

Now we show that $q^{(\infty)}$ is the largest solution in $[0,1]^K$.
Suppose $\hat q=(\hat q_1, \hat q_2, \ldots, \hat q_K)$ is another solution  in $[0,1]^K$.
We first show by induction that $\hat q \le q^{(i)}$ for all $i$.
Clearly, $\hat q_k \le 1 =q_k^{(0)}$ for all $k$.
Assume that $\hat q \le q^{(i-1)}$ as the induction hypothesis.
As $\hat q$ is a solution of \req{stable0000}, we have from the induction hypothesis, the monotonicity of $P_{{\rm suc},k,j}(\rho)$, and
\req{tag6666dmulm} that
\bear{stable3333b}
&&\hat q_{k}\nonumber\\
&&=\lambda_{k}\Big (1- \sum_{j=1}^J r_{k,j} P_{{\rm suc},k,j}(\hat q \circ G \circ \Lambda^\prime(1) \circ \Rj)\Big )\nonumber\\
&&\le \lambda_{k}\Big (1- \sum_{j=1}^J r_{k,j} P_{{\rm suc},k,j}(q^{(i-1)}\circ G \circ \Lambda^\prime(1) \circ \Rj)\Big )\nonumber\\
&&=q_k^{(i)}
\eear
This completes the induction for $\hat q \le q^{(i)}$ for all $i$.
Letting $i \to\infty$ yields that
$\hat q_{k} \le q_k^{(\infty)}$ for all $k$.
\eproof

\bproof
(\rthe{stable})

\noindent {\em (if part)}
If ${\bf 0}$ is the unique solution of the $K$ equations in \req{stable0000}, it then follows from \rlem{limit} that $q^{(\infty)}={\bf 0}$.
In view of $P_{{\rm suc},k,j}({\bf 0})=1$, $\Lambda_k(0)=0$, and the continuity of $P_{{\rm suc},k,j}(\rho)$, we have from \req{mean8888thumulam} that
\beq{srates1111a}
\lim_{i \to\infty} \tilde P_{{\rm suc},k}^{(i)}(G)=1, \quad k=1,2,\ldots, K.
\eeq

\noindent {\em (only if part)}

If $G$ is stable, then we have from \req{mean8888thumulam} that
\bear{srates1111b}
&&1=\lim_{i \to\infty} \tilde P_{{\rm suc},k}^{(i)}(G) \nonumber\\
 &&=1-\Lambda_k \Big  (1- \sum_{j=1}^J r_{k,j} P_{{\rm suc},k,j}(q^{(\infty)} \circ G \circ \Lambda^\prime(1) \circ \Rj)\Big).\nonumber \\
\eear
Since $\Lambda_k(x)=0$ if and  only if $x=0$, we then have for all $k$,
\beq{srates1135}
\sum_{j=1}^J r_{k,j} P_{{\rm suc},k,j}(q^{(\infty)} \circ G \circ \Lambda^\prime(1) \circ \Rj)=1.
\eeq
As $\sum_{j=1}^J r_{k,j}=1$, we have
\beq{srates1136}
\sum_{j=1}^J r_{k,j}(1- P_{{\rm suc},k,j}(q^{(\infty)} \circ G \circ \Lambda^\prime(1) \circ \Rj))=0.
\eeq
Under (A3), we know that $r_{k,j} >0$ for all $k$ and $j$. This then leads to
\beq{srates1137}
 P_{{\rm suc},k,j}(q^{(\infty)} \circ G \circ \Lambda^\prime(1) \circ \Rj)=1,
\eeq
for all $k$ and $j$.
Under (A2), we know that $P_{{\rm suc},k,j}({\bf \rho})=1$ if and only if  ${\bf \rho} = {\bf 0}$. Thus,
 \beq{srates1139}
 q^{(\infty)} \circ G \circ \Lambda^\prime(1) \circ \Rj={\bf 0}.
\eeq
Since $r_{k,j} >0$, all the elements in $\Rj$ are positive. Also, as $\Lambda_k^\prime(1)$ is the average number of copies of class $k$ packets, all the elements in  $\Lambda^\prime(1)$ are positive. For the positive offered load $G$, all the elements in $G$ are positive. It then follows from \req{srates1139}
that $q^{(\infty)}={\bf 0}$. As $q^{(\infty)}$ is the largest solution in $[0,1]^K$ of the  $K$ fixed point equations in \req{stable0000} (from \rlem{limit}), we conclude that $q^{(\infty)}={\bf 0}$ is the {\em unique} solution in $[0,1]^K$.
\eproof

\bsubsec{Existence of the stability region}{existence}

In the following theorem, we show the existence of stability regions for coded Poisson receivers.

\bthe{region}
Suppose that (A1), (A2), and (A3) hold.
If a positive Poisson offered load $\hat G=(\hat G_1, \hat G_2, \ldots, \hat G_K)$ (with $\hat G_k >0$ for all $k$) is stable, then
any Poisson offered load $G$ with $G \le \hat G$ is also stable.
\ethe

\bproof
Let $q^{(i)}(G)$ be the probability $q^{(i)}$ when the system of coded Poisson receivers is subject to the Poisson offered load $G$.
We will show by induction that $q^{(i)}(G) \le  q^{(i)}(\hat G)$ if $G \le \hat G$.
For $i=0$, we have $q^{(0)}_k (G)=q^{(0)}_k(\hat G)=1$ for all $k$.
Now assume that  $q^{(i-1)}(G) \le  q^{(i-1)}(\hat G)$ as the induction hypothesis.
It then follows from \req{tag6666dmulm}, the assumption that  $P_{{\rm suc},k,j}(\rho)$ is  decreasing in $\rho$ in (A1), and the
induction hypothesis
that
\bear{stable3333r}
&&q_{k}^{(i)}(G)\nonumber\\
&&=\lambda_{k}\Big (1- \sum_{j=1}^J r_{k,j} P_{{\rm suc},k,j}(q^{(i-1)}(G) \circ G \circ \Lambda^\prime(1) \circ \Rj)\Big )\nonumber\\
&&\le \lambda_{k}\Big (1- \sum_{j=1}^J r_{k,j} P_{{\rm suc},k,j}(q^{(i-1)}(\hat G) \circ \hat G \circ \Lambda^\prime(1) \circ \Rj)\Big )\nonumber\\
&&=q_{k}^{(i)}(\hat G).
\eear
This completes the induction for $q^{(i)}(G) \le  q^{(i)}(\hat G)$.
Letting $i \to \infty$ yields $q^{(\infty)}(G) \le  q^{(\infty)}(\hat G)$.
Since $\hat G$ is stable, we have from \rthe{stable} that $q^{(\infty)}(\hat G)={\bf 0}$.
This then implies that $q^{(\infty)}(G)={\bf 0}$ and thus $G$ is also stable.
\eproof

The result in \rthe{region} enables us to define the stability region of a system of coded Poisson receivers.

\bdefin{region}
Under (A1), (A2), and (A3), the stability region $S$ is defined as the maximal stable set such that
(i) any $G \in S$ is stable, and (ii) any $G \not \in S$ is not stable.
\edefin

\bsubsec{Characterization of the stability region}{characterization}

One important question is to characterize the stability region of a system of coded Poisson receivers.
For this, we provide in the following lemma a sufficient condition for
$\bf 0$ to be the unique solution of \req{stable0000}.
\bthe{stableq}
Suppose that (A1) holds.
If for all $q=(q_1, q_2,\ldots,q_K) \in [0,1]^K$ and $q \ne 0$, there exists some $k \in \{1,2,\ldots, K\}$ such that
\beq{stable1111}
q_{k}>\lambda_{k}\Big (1- \sum_{j=1}^J r_{k,j} P_{{\rm suc},k,j}(q \circ G \circ \Lambda^\prime(1) \circ \Rj)\Big ).
\eeq
Then $\bf 0$ is the unique solution of \req{stable0000} and the offered load $G$ is thus stable.
\ethe

\bproof
From \rlem{limit}, we know that $q^{(\infty)}$ is the largest solution  of the  $K$ fixed point equations in \req{stable0000} (among all the solutions in $[0,1]^K$).
Thus, if $q^{(\infty)}={\bf 0}$, then ${\bf 0}$ is the unique solution.
In view of \rthe{stable},
it suffices to show that $q^{(\infty)}={\bf 0}$.
The intuition of the condition in \req{stable1111} is that $q^{(\infty)}$ will continue to decrease if it is not ${\bf 0}$.
We prove this by contradiction.
Suppose that $q^{(\infty)}\ne {\bf 0}$. Then
we have from \rlem{limit} and \req{stable1111}
that
\bear{stable5555c}
&&q_{k}^{(\infty)} \nonumber\\
&&=\lambda_{k}\Big (1- \sum_{j=1}^J r_{k,j} P_{{\rm suc},k,j}(q^{(\infty)} \circ G \circ \Lambda^\prime(1) \circ \Rj)\Big )\nonumber\\
&&<q_{k}^{(\infty)},
\eear
for some $k \in \{1,2,\ldots, K\}$.
Thus, we reach a contraction and  $q^{(\infty)}={\bf 0}$.
\eproof

Instead of using the limiting vector of $q^{(i)}$, one can use
the limiting vector of $p^{(i)}$  to show the stability of a system of coded Poisson receivers.
This is shown in the following corollary that requires every packet in the system of the coded Poisson receivers to be transmitted at least twice.

\bcor{stablep}
Suppose that (A1) holds and  every packet in the system of the coded Poisson receivers is transmitted at least twice, i.e.,
\beq{stablep1111}\Lambda_{k,1}=0
\eeq
 for all $k=1,2,\ldots,K$.
If for all $p=(p_1, p_2, \ldots, p_K)$ in $[0,1]^K$ and $p \ne {\bf 0}$,
there exists some $k \in\{1,2, \ldots, K\}$ such that
\beq{stable1111p}
p_{k}>1- \sum_{j=1}^J r_{k,j} P_{{\rm suc},k,j}( G \circ \Lambda^\prime(p) \circ \Rj).
\eeq
Then $\bf 0$ is the unique solution of \req{stable0000} and the offered load $G$ is thus stable.
\ecor

\bproof
We will show that the condition in \req{stable1111p}  implies the condition in \req{stable1111} of \rthe{stableq} under the additional assumption that every packet in the system of the coded Poisson receivers is transmitted at least twice.
Since $\Lambda_{k,1}=0$ for all $k$ in \req{stablep1111}, we have from \req{mean2222mulr} that $\lambda_{k,0}=0$ for all $k$.
In view of \req{mean3333mulr}, we know that $\lambda_k(0)=0$, $\lambda_k(1)=1$, and $\lambda_k(x)$  is strictly increasing in $x$ for $x \in [0,1]$. As such,  the inverse function of $\lambda_k(x)$ exists for $x \in [0,1]$. Moreover, the inverse function is also strictly increasing. Let $\lambda_k^{-1}(\cdot)$ be the inverse function of $\lambda_k(x)$ and $p_k=\lambda_k^{-1}(q_k)$. Note from \req{mean3344mulr} that
\beq{equiv2222}
q_k \Lambda_k^\prime(1)= \lambda_k(p_k) \Lambda_k^\prime(1)= \Lambda_k^\prime(p_k),
\eeq
and thus
\beq{equiv3333}
q \circ G \circ \Lambda^\prime(1) \circ \Rj =G \circ \Lambda^\prime(p) \circ \Rj.
\eeq
Taking the function $\lambda_k(\cdot)$ on both sides of \req{stable1111p} yields \req{stable1111}.
\eproof

\bex{irsa}{(Irregular Repetition Slotted
ALOHA (IRSA) \cite{liva2011graph})}
IRSA can be considered as a special case of the coded Poisson receivers with a single class of users ($K=1$), a single class of receivers ($J=1$), and the success probability function $P_{{\rm suc}}(\rho)=e^{-\rho}$ (see \rex{TfoldALOHA} for $D=1$). For this special case, we assume that $r_{11}=1$ and $F_1=1$.
Suppose that the number of copies of a packet in IRSA is at least 2 and it has the following distribution:
\beq{extm1111irsa}
P(L=k+1)=\Lambda_{k+1}, \;k=1,\dots, K.
\eeq

From \rcor{stablep},
the sufficient condition in \req{stable1111p}  is
\beq{stable5555irsa}
p>1-  \exp(- G \Lambda^\prime(p))
\eeq
for $0< p \le 1$.
Thus, $G$ is stable for IRSA if $G < G^*$, where
\beq{stable6666irsa}
G^*=\inf_{0 <p \le 1}\frac{-\log(1-p)}{\Lambda^\prime(p)}.
\eeq
Note that
\bear{stable7777irsa}
-\log(1-p)&=&p+\frac{(p)^2}{2}+\frac{(p)^3}{3}+ \cdots \nonumber\\
&=&\sum_{k=1}^\infty \frac{(p)^k}{k},
\eear
and
that
\beq{stable4444irsa}
\Lambda^\prime(p)=\sum_{k=1}^K  (k+1)\Lambda_{k+1}p^k.
\eeq
In view of \req{stable4444irsa}, if we choose
\beq{stable8888irsa}
\Lambda_{k+1}=\frac{1}{k (k+1)}
\eeq
for all $k$, then
\beq{stable9999irsa}
\Lambda^\prime(p)\approx -\log(1-p)
\eeq
and $G^* \approx 1$
when $K$ is large. This is exactly the choice of the degree distribution in \cite{narayanan2012iterative} that can achieve throughput efficiency arbitrarily close to 1 in IRSA.
\eex

\bsubsec{Weak stability}{wstability}

The notion of stability might be too strong for practical communication channels as one cannot guarantee the success of a packet transmission due to noise, channel fading, and decoding errors.
For this, we relax the assumption (A1) to (A1w) to generalize the notion of stability to the notion of
weak stability below.

\begin{description}
\item[(A1w)] For all $k=1, \ldots, K$ and $j=1,2, \ldots, J$, the success  probability function $P_{{\rm suc},k,j}(\rho)$ is  a continuous and decreasing function of $\rho$.
\end{description}

We note that \rlem{limit} still holds under (A1w) and we can define a weaker notion of stability by using $q^{(\infty)}(G)$ in \req{stable4444}. As shown   in \rthe{stable}, the necessary and sufficient condition for stability is when ${\bf 0}$
is the unique solution of the  $K$ equations in
\req{stable0000}. Motivated by this, we propose the following definition
of weak stability.

\bdefin{weaks}{(\bf weak stability and weak stability region of coded Poisson receivers)}
Suppose that (A1w) holds for the coded Poisson receiver described in \rsec{mulr}.
A Poisson offered load $\hat G=(\hat G_1, \hat G_2, \ldots, \hat G_K)$ to the coded Poisson receiver  is said to be {\em weakly stable}  if for any $G \le \hat G$, $q^{(\infty)}(G)$ is  the unique solution  in $[0,1]^K$ of the  $K$ equations in
\req{stable0000}.
Moreover,
the weak stability region $S$ is defined as the maximal weakly  stable set such that
(i) any $G \in S$ is weakly stable, and (ii) any $G \not \in S$ is not weakly stable.
\edefin

It is known from \rlem{limit} that $q^{(\infty)}(G)$ is the largest solution of \req{stable0000} among all the solutions in $[0,1]^K$.
Note that $q^{(\infty)}(G)$ is obtained from the initial vector $q^{(0)}=(1,1,\ldots,1)$. If we start from the initial vector $q^{(0)}={\bf 0}$, then $q^{(i)}$ is an increasing sequence, and the limit obtained this way, denoted by $\qinf(G)$, is the smallest solution of \req{stable0000} among all the solutions in $[0,1]^K$.
This leads to the following necessary and sufficient condition for weak stability.

\bprop{weakcon}
Suppose that (A1w) hold.
A Poisson offered load $\hat G=(\hat G_1, \hat G_2, \ldots, \hat G_K)$ to the coded Poisson receiver  is  weakly stable  if and only if $q^{(\infty)}(G)=\qinf(G)$ for any $G \le \hat G$.
\eprop

In general, due to the nonlinearity of the success probability function, it is very difficult to characterize the weak stability region analytically. The result in \rprop{weakcon}
  provides a numerical method to identify the weak stability region. To see the intuition of the result in \rprop{weakcon}, let us consider the case with $K=1$ and let $G^*=\inf\{G: q^{(\infty)}(G)\ne \qinf(G)\}$
  be the smallest offered load $G$ such that there are at least two solutions in
\req{stable0000}. Then the function $q^{(\infty)}(G)$ is not continuous at $G^*$ as
$$q^{(\infty)}(G^*-\delta)= \qinf(G^*-\delta) \le \qinf(G^*)<
q^{(\infty)}(G^*)$$
for any $\delta>0$. Thus, $G^*$ is a percolation threshold and the weak stability region is
$\{G: G < G^*\}$.  In \rsec{num}, we will use this method  to identify the weak stability region for a CPR system with decoding errors.

\bsubsec{$\epsilon$-stability}{estability}

The notion of weak stability does not provide a specific guarantee for the probability that a packet is successfully received. To provide guarantees, we add one more assumption below.

\begin{description}
\item[(A1g)]
For a $K$-vector $\epsilon=(\epsilon_1, \epsilon_2, \ldots, \epsilon_K)$ in $[0,1]^K$, there exists a nonempty set $\Gamma_0(\epsilon)$ such that for all $\rho \in \Gamma_0(\epsilon)$, $j=1,2, \ldots, J$, and $k=1,2, \ldots, K$,
    $$P_{{\rm suc},k,j}(\rho)\ge 1-\epsilon_k.$$
\end{description}

The assumption  (A1g)  reduces to the assumption $P_{{\rm suc},k,j}({\bf 0})=1$ in (A1) if $\epsilon={\bf 0}$ and the nonempty set $\Gamma_0(\epsilon)$ only contains the point ${\bf 0}$.

\bdefin{wsrates}{(\bf $\epsilon$-stability of coded Poisson receivers)}
Consider the coded Poisson receiver described in \rsec{mulr}.
A Poisson offered load $G=(G_1, G_2, \ldots, G_K)$ to the coded Poisson receiver  is said to be {\em $\epsilon$-stable}  if $q^{(\infty)}$ is in $\Gamma_1(\epsilon)$,
where
\beq{r1epsilon}
\Gamma_1(\epsilon)=[0, \lambda_1(\epsilon_1)] \times [0, \lambda_2(\epsilon_2)] \times \ldots \times  [0, \lambda_K(\epsilon_K)].
\eeq
\edefin

In the following corollary, we show that the probability of a class $k$ packet is successfully received is at least
$1-\Lambda_k (\epsilon_k)$ for a $\epsilon$-stable CPR.

\bthe{wstable}
Suppose that (A1w) and (A1g) hold and that every packet in the system of the coded Poisson receivers is transmitted at least twice, i.e.,
\beq{stablep1111w}\Lambda_{k,1}=0.
\eeq
If a Poisson offered load $G$  is $\epsilon$-stable, then for $k=1,2,\ldots, K$,
\beq{wsrates1111}
\lim_{i \to\infty} \tilde P_{{\rm suc},k}^{(i)}(G)\ge 1-\Lambda_k (\epsilon_k),
\eeq
where $\tilde P_{{\rm suc},k}^{(i)}(G)$ is defined in \req{mean8888thumulam}.
\ethe

\bproof
As shown in \rcor{stablep},
the inverse function of $\lambda_k(x)$ exists when every packet in the system of the coded Poisson receivers is transmitted at least twice.
 Since $q^{(\infty)}$ is a solution of \req{stable0000} in \rlem{limit},
we have
\beq{weaks1111}
\lambda_k^{-1}(q_k^{(\infty)})=1- \sum_{j=1}^J r_{k,j} P_{{\rm suc},k,j}(q^{(\infty)} \circ G \circ \Lambda^\prime(1) \circ \Rj).
\eeq
Note from \req{mean8888thumulam}, the continuity of $P_{{\rm suc},k,j}(\rho)$, and \req{weaks1111} that
\bear{weaks2222}
&&\lim_{i \to\infty} \tilde P_{{\rm suc},k}^{(i)}(G)\nonumber\\
&&=1-\Lambda_k \Big  (1- \sum_{j=1}^J r_{k,j} P_{{\rm suc},k,j}(q^{(\infty)} \circ G \circ \Lambda^\prime(1) \circ \Rj)\Big)\nonumber\\
&&=1-\Lambda_k  (\lambda_k^{-1}(q_k^{(\infty)})).
\eear
Since $q^{(\infty)}$ is in $\Gamma_1(\epsilon)$, we then have from the monotonicity of the function $\Lambda_k(\cdot)$ that
$$\lim_{i \to\infty} \tilde P_{{\rm suc},k}^{(i)}(G)\ge 1-\Lambda_k  (\lambda_k^{-1}(\lambda_k(\epsilon_k)))
=1-\Lambda_k(\epsilon_k).
$$
\eproof




In the following corollary, we show the existence of $\epsilon$-stability regions for coded Poisson receivers.

\bcor{wregion}
Suppose that (A1w) and (A1g) hold.
If a Poisson offered load $\hat G=(\hat G_1, \hat G_2, \ldots, \hat G_K)$
is $\epsilon$-stable, then
any Poisson offered load $G$ with $G \le \hat G$ is also $\epsilon$-stable.
\ecor

\bproof
The proof is basically the same as that in \rthe{region}.
Let $q^{(\infty)}(G)$ be the probability $q^{(\infty)}$ when the system of coded Poisson receivers is subject to the Poisson offered load $G$.
It is shown in \rthe{region} that $q^{(\infty)}(G) \le  q^{(\infty)}(\hat G)$ if $G \le \hat G$.
Since $\hat G$ is $\epsilon$-stable, $ q^{(\infty)}(\hat G)$ is in $\Gamma_1(\epsilon)$.
Thus, for any $G \le \hat G$, we know that $q^{(\infty)}(G)$ is also in $\Gamma_1(\epsilon)$ and thus $\epsilon$-stable.
\eproof

The result in \rcor{wregion} enables us to define the $\epsilon$-stability region of a system of coded Poisson receivers.

\bdefin{wregion}
Under (A1w) and (A1g), the $\epsilon$-stability region $S(\epsilon)$ is defined as the maximal $\epsilon$-stable set such that
(i) any $G \in S(\epsilon)$ is $\epsilon$-stable, and (ii) any $G \not \in S(\epsilon)$ is not $\epsilon$-stable.
\edefin

Note from (A1g) that $S(\epsilon)$ is a nonempty set as ${\bf 0}$ is in $S(\epsilon)$.


Analogous to \rthe{stableq}, we provide  a sufficient condition to characterize
the $\epsilon$-stability region of a system of coded Poisson receivers.

\bcor{wstableq}
Suppose that (A1w) and (A1g) hold.
If for all $q=(q_1, q_2,\ldots,q_K) \in [0,1]^K$ and $q \not \in \Gamma_1(\epsilon)$, there exists some $k$ such that
\beq{wstable1111}
q_{k}>\lambda_{k}\Big (1- \sum_{j=1}^J r_{k,j} P_{{\rm suc},k,j}(q \circ G \circ \Lambda^\prime(1) \circ \Rj)\Big ).
\eeq
Then  the offered load $G$ is $\epsilon$-stable.
\ecor

\bproof
Analogous to the proof of \rthe{stableq},
we prove this by contradiction.
Suppose that $q^{(\infty)}\not \in \Gamma_1(\epsilon)$. Then
we have from \rlem{limit} and \req{wstable1111}
that
\bear{wstable5555c}
&&q_{k}^{(\infty)} \nonumber\\
&&=\lambda_{k}\Big (1- \sum_{j=1}^J r_{k,j} P_{{\rm suc},k,j}(q^{(\infty)} \circ G \circ \Lambda^\prime(1) \circ \Rj)\Big )\nonumber\\
&&<q_{k}^{(\infty)},
\eear
for some $k \in \{1,2,\ldots, K\}$.
Thus, we reach a contraction and  $q^{(\infty)}$ must be in $\Gamma_1(\epsilon)$.
\eproof

As in \rcor{stablep}, one can use
the limiting vector of $p^{(i)}$  to show the $\epsilon$-stability of a system of coded Poisson receivers.
The proof for \rcor{stablepw} is the same as \rcor{stablep} and is thus omitted.

\bcor{stablepw}
Suppose that (A1w) and (A1g) hold and that every packet in the system of the coded Poisson receivers is transmitted at least twice.
Consider the bounded region
\beq{r2epsilon}
\Gamma_2(\epsilon)=[0, \epsilon_1] \times [0,\epsilon_2]\times \ldots \times [0,\epsilon_K].
\eeq
If for all $p=(p_1, p_2, \ldots, p_K)$ in $[0,1]^K$ and $p \not \in \Gamma_2(\epsilon)$,
there exists some $k \in\{1,2, \ldots, K\}$ such that
\beq{stable1111pw}
p_{k}>1- \sum_{j=1}^J r_{k,j} P_{{\rm suc},k,j}( G \circ \Lambda^\prime(p) \circ \Rj).
\eeq
Then the offered load $G$ is thus $\epsilon$-stable.
\ecor

\section{Numerical Results}
\label{sec:num}

\subsection{IRSA with two classes of users and two classes of receivers}
\label{sec:irsa2}

\subsubsection{Stability region}
\label{sec:stability1}

In this section, consider a special case of the system of coded Poisson receivers with two classes of users ($K=2$), two classes of receivers ($J=2$), and the success probability function $P_{{\rm suc}}(\rho)=e^{-\rho}$.
As the success probability function $P_{{\rm suc}}(\rho)=e^{-\rho}$ is taken from the slotted ALOHA (\rex{TfoldALOHA} for $D=1$), we call such a system the {\em IRSA system with two classes of users and two classes of receivers}.
Clearly, the condition (A1) in \rcor{stablep} is satisfied.
Let  $\Lambda_1(x)$ and $\Lambda_2(x)$ be the generating functions of the degree distributions of the two classes of users.
From \rcor{stablep}, a Poisson offered load $G=(G_1,G_2)$ to such a system is stable
if  at least one of the following two inequalities is satisfied when $(p_1,p_2)\ne (0,0)$:
\bear{irsa1111a}
&&p_{1}>1-  r_{1,1} e^{-G_1 \Lambda^\prime_1(p_1)r_{1,1}/F_1-G_2 \Lambda^\prime_2(p_2)r_{2,1}/F_1}\nonumber\\
&&\quad-r_{1,2} e^{-G_1 \Lambda^\prime_1(p_1)r_{1,2}/F_2-G_2 \Lambda^\prime_2(p_2)r_{2,2}/F_2},
\eear
\bear{irsa1111b}
&&p_{2}>1-  r_{2,1} e^{-G_1 \Lambda^\prime_1(p_1)r_{1,1}/F_1-G_2 \Lambda^\prime_2(p_2)r_{2,1}/F_1}\nonumber\\
&&\quad-r_{2,2} e^{-G_1 \Lambda^\prime_1(p_1)r_{1,2}/F_2-G_2 \Lambda^\prime_2(p_2)r_{2,2}/F_2} .
\eear

Motivated by the circuit routing policies in circuit-switching networks (see, e.g., \cite{kelly1991loss,chang1997large}),
 we consider the following four packet routing policies:
 \begin{description}
  \item[(i)] Complete sharing:  every packet has an equal probability to be routed to the two classes of receivers, i.e., $r_{11}=r_{22}=r_{12}=r_{21}=0.5$.

   \item[(ii)] Receiver reservation: class 1 receivers are reserved for class 1 packets only. On the other hand, class 2 receivers are shared by these two classes of users. Specifically, class 1 packets are routed to the two classes of receivers with an equal probability, i.e., $r_{11}=r_{12}=0.5$, and class 2 packets are routed to the class 2 receivers, i.e., $r_{21}=0$, $r_{22}=1$.
   \item[(iii)] Nearly complete partitioning: class 1 receivers are reserved for class 1 packets only, and class 1 packets are routed to class 1 (resp. 2) receivers with probability 0.9 (resp. 0.1), i.e., $r_{11}=0.9$, $r_{12}=0.1$, $r_{21}=0$, $r_{22}=1$.

 \item[(iv)] Nonuniform sharing:  class 1 (resp. 2) packets are routed to class 1 (resp. 2) receivers
   with probability 0.3, and  routed to class 2 (resp. 1) with probability 0.7, i.e., $r_{11}=r_{22}=0.3$, and $r_{12}=r_{21}=0.7$.

   \end{description}

\begin{figure}[ht]
	\centering
	\includegraphics[width=0.43\textwidth]{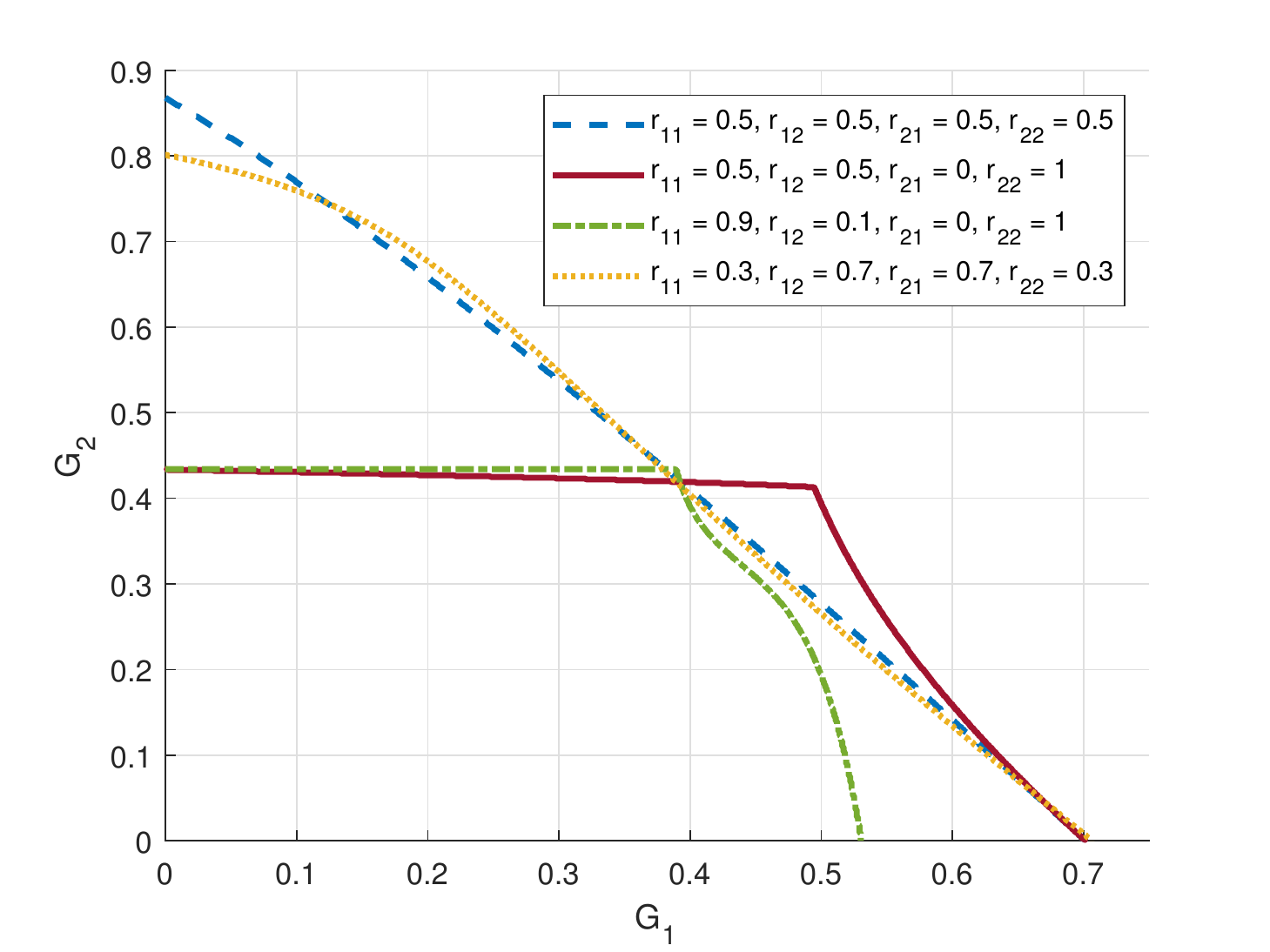}
	\caption{The stability region for the IRSA system with two classes of users and two classes of receivers under four different routing policies: (i) complete sharing ($r_{11}=r_{22}=r_{12}=r_{21}=0.5$) marked with the blue curve, (ii) receiver reservation ($r_{11}=r_{12}=0.5$, $r_{21}=0$, $r_{22}=1$) marked with the red curve, (iii) nearly complete partitioning ($r_{11}=0.9$, $r_{12}=0.1$, $r_{21}=0$, $r_{22}=1$) marked with the green curve, and (iv) nonuniform sharing ($r_{11}=r_{22}=0.3$, $r_{12}=r_{21}=0.7$) marked with the yellow curve.
}
	\label{fig:stableregion}
\end{figure}

 For our experiments, we set the numbers of receivers of these two classes of receivers to be the same, i.e., $F_1=F_2=0.5$. Also, we choose $\Lambda_1(x)=x^5$ and $\Lambda_2(x)=0.5102x^2+0.4898x^4$, where $\Lambda_2(x)$ is taken  from Table 1  of \cite{liva2011graph} for achieving a large percolation threshold of 0.868 in IRSA with a single class of users.
We stop the computation for $q^{(\infty)}$ when the number of SIC iterations reaches $500$ ($i=500$).
In \rfig{stableregion},
we show the stability regions of these four policies (marked with  four different colored curves) obtained from the grid search with the step size of $0.001$ for both $G_1$ and $G_2$. For the sake of numerical stability in our computation, we round up $\tilde P_{{\rm suc,k}}^{(i)}(G_1,G_2)$ to 1 if its computed value is larger than $0.99999$.

\noindent {\bf (i) Complete sharing:}

As shown in \rfig{stableregion}, the stability region of the complete sharing policy is separated by the blue curve. Since the routing probabilities of each packet to the two classes of receivers are the same, there is no difference for the two classes of packets at the receivers. In other words, the complete sharing policy can be viewed as an IRSA with a {\em single} class of users with the degree distribution
$$\Lambda_1(x)\frac{G_1}{G_1+G_2}+\Lambda_2(x)\frac{G_2}{G_1+G_2}.$$
In view of \req{stable5555irsa}, a sufficient condition for $(G_1, G_2)$ to be stable is
$$p>1- \exp(- (G_1+G_2) (\Lambda_1^\prime(p)\frac{G_1}{G_1+G_2}+\Lambda_2^\prime (p)\frac{G_2}{G_1+G_2}))
$$
for all $0< p \le 1$.
In view of the sufficient condition, it is easy to see that the stability region of  the complete sharing policy
is convex, i.e., if $(G_1^\prime,G_2^\prime)$ and $(G_1^{\prime\prime}, G_2^{\prime\prime})$ are stable, then
$(\alpha G_1^\prime+(1-\alpha)G_1^{\prime\prime}, \alpha G_2^\prime+(1-\alpha)G_2^{\prime\prime})$ is also stable for any $0 \le \alpha\le 1$. We also note that the blue curve in \rfig{stableregion} is not a straight line at a higher resolution of the plot.

\noindent {\bf (ii) Receiver reservation:}

The stability region for this policy is shown in the red curve in \rfig{stableregion}. The stability region of this policy is no longer convex. For this policy,  the maximum stable  load for class 2 users is roughly 0.43, which is considerably smaller than 0.868 of the complete sharing policy. This is because class 2 users are not allowed to access class 1 receivers. However, when $G_2 < 0.43$,  the  maximum stable  load for class 1 users can be larger than that of the complete sharing policy.
The corner point of the red curve is roughly $(G_1,G_2)=(0.494,0.413)$.  This is due to the reservation policy that reserves class 1 receivers to class 1 users. To see the insight of this, note that when the traffic loads of these two classes of users are very high, the system is highly congested and the probability of receiving a  packet successfully in slotted ALOHA is very low. In such a setting, it is preferable to limit class 2 users from entering a certain portion of the receivers, i.e., class 1 receivers in this reservation policy, to lighten the load in that portion of receivers. By  doing so, part of class 1 packets can still get through in this portion of receivers. The coded random access scheme with SIC can then further remove the other copies of the successfully received packets from the rest of receivers.

\noindent {\bf (iii) Nearly complete partitioning:}

The stability region for this policy is shown in the green curve in \rfig{stableregion}. The stability region of this policy is also not convex, and it is almost included in the stability region of the complete sharing policy and that of the receiver reservation policy. This shows that partitioning is not a good policy as the resources (receivers) are neither shared nor reserved.

\noindent {\bf (iv) Nonuniform sharing:}

The stability region for this policy is shown in the yellow curve in \rfig{stableregion}. We note that nonuniform sharing can alter the shape of the stability region and it is different from that of the complete sharing policy.
Even though nonuniform sharing can increase the stability for certain offered loads (i.e., some part of the yellow curve is above the blue curve), the gain is rather insignificant. It seems that the complete sharing policy is a better choice for achieving the stability of these two classes of users.

\subsubsection{Throughput}
\label{sec:throughput}

In this section, we discuss the throughput of the IRSA system with two classes of users. For this, we
consider the receiver reservation policy with $r_{11}=0.5$, $r_{12}=0.5$, $r_{21}=0$, $r_{22}=1$.
In this setting, we still choose $F_1=F_2=0.5$.  Also, we use the same degree distributions as in the previous section, i.e.,  $\Lambda_1(x)=x^5$ and $\Lambda_2(x)=0.5102x^2+0.4898x^4$. From \req{mean8888thumulam}, we can compute the  success probability function for class $1$ (resp. class $2$) users, i.e., $\tilde P_{{\rm suc},1}^{(\infty)}(G)$ (resp. $\tilde P_{{\rm suc},2}^{(\infty)}(G)$).
Then  the total throughput  of these two classes users is
$$G_1\tilde P_{{\rm suc},1}^{(\infty)}(G)+G_2\tilde P_{{\rm suc},2}^{(\infty)}(G).$$

\begin{figure}[ht]
	\centering
	\includegraphics[width=0.43\textwidth]{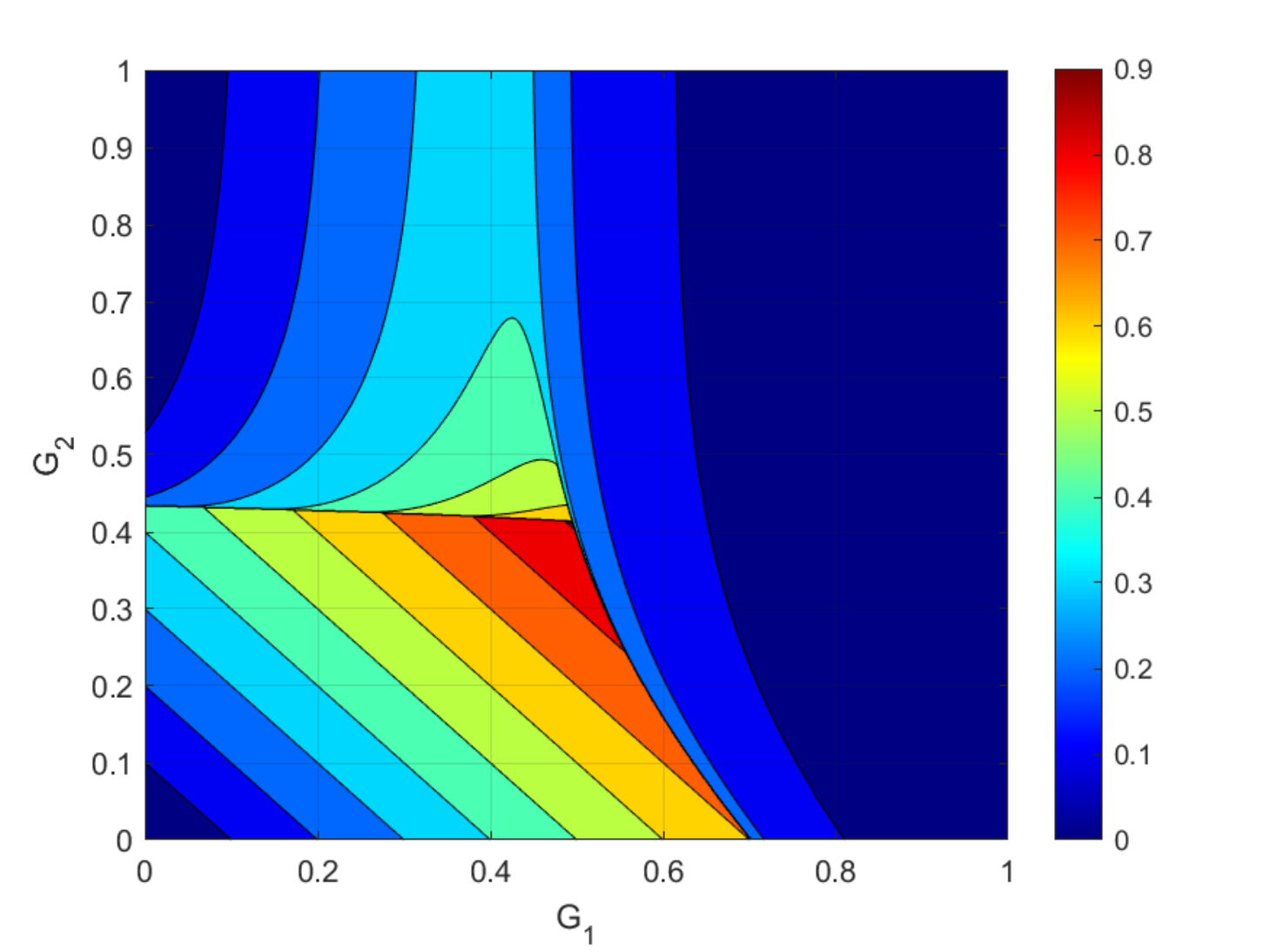}
	\caption{The total throughput of the two classes of users as a function of $G=(G_1,G_2)$ for the receiver reservation policy with the routing probabilities $r_{11}=0.5$, $r_{12}=0.5$, $r_{21}=0$, $r_{22}=1$. The degree distributions of the two classes of users are $\Lambda_1(x)=x^5$ and $\Lambda_2(x)=0.5102x^2+0.4898x^4$.}
	\label{fig:throall}
\end{figure}

In \rfig{throall}, we show the total throughput as a function of $G=(G_1, G_2)$. Basically, it can be partitioned into three regions:

\noindent
(i) both classes of users are stable: when $G_1<0.494$ and $G_2<0.413$, the total throughput is increasing with respect to the total offered load $G_1+G_2$ as both classes are stable. This region is marked with various strips of different colors that show the linear increase of the total throughput. The maximum total throughput is roughly 0.9 (the red part in \rfig{throall}).

\noindent (ii) both classes of users are not stable: this region is marked with the blue color and the dark blue color. In this region, both classes of users are not stable and the total throughput decreases rapidly to 0 with respect to $G_1+G_2$.

\noindent (iii) class 1 users are stable but class 2 users are not stable: this is the region corresponding to $G_2>0.413$ and $G_1<0.494$. The total throughput decreases with respect to the increase of $G_2$ and the maximum total throughput in this region is roughly $0.6$ (the part marked in yellow in \rfig{throall} near $G_1=0.494$ and $G_2=0.413$).

\subsubsection{weak stability and $\epsilon$-stability}
\label{sec:weaknum}

In this section, we consider the effect of decoding errors in the IRSA system.
As discussed in \rex{errorALOHA}, we let $p_{\rm err}$ be the probability of decoding errors, and thus each
Poisson receiver has the success probability function $P_{{\rm suc}}(\rho)=(1-p_{\rm err})e^{-\rho}$.

\begin{figure}[tb]
	\begin{center}
		\begin{tabular}{p{0.23\textwidth}p{0.23\textwidth}}
			\includegraphics[width=0.23\textwidth]{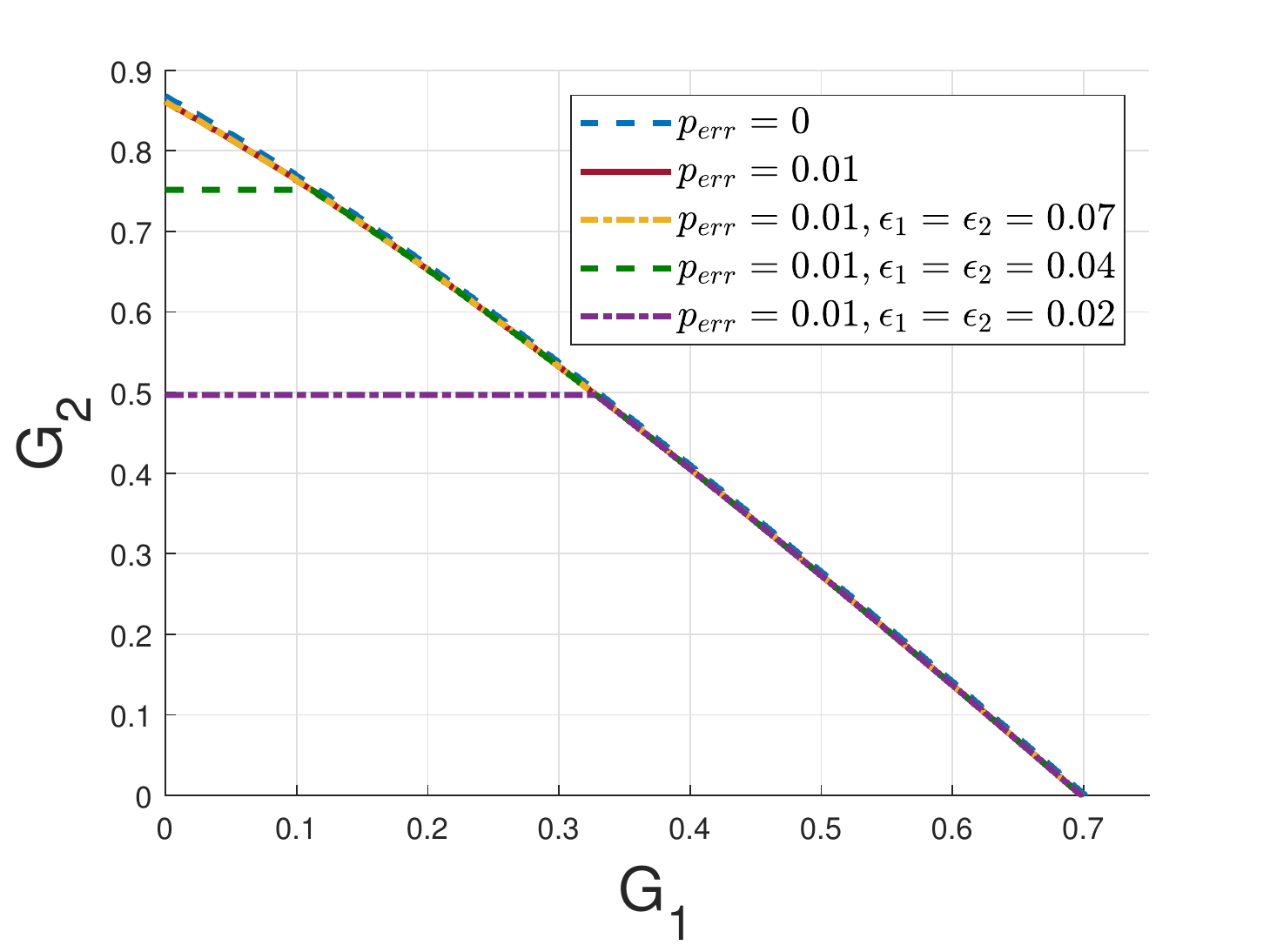} &
			\includegraphics[width=0.23\textwidth]{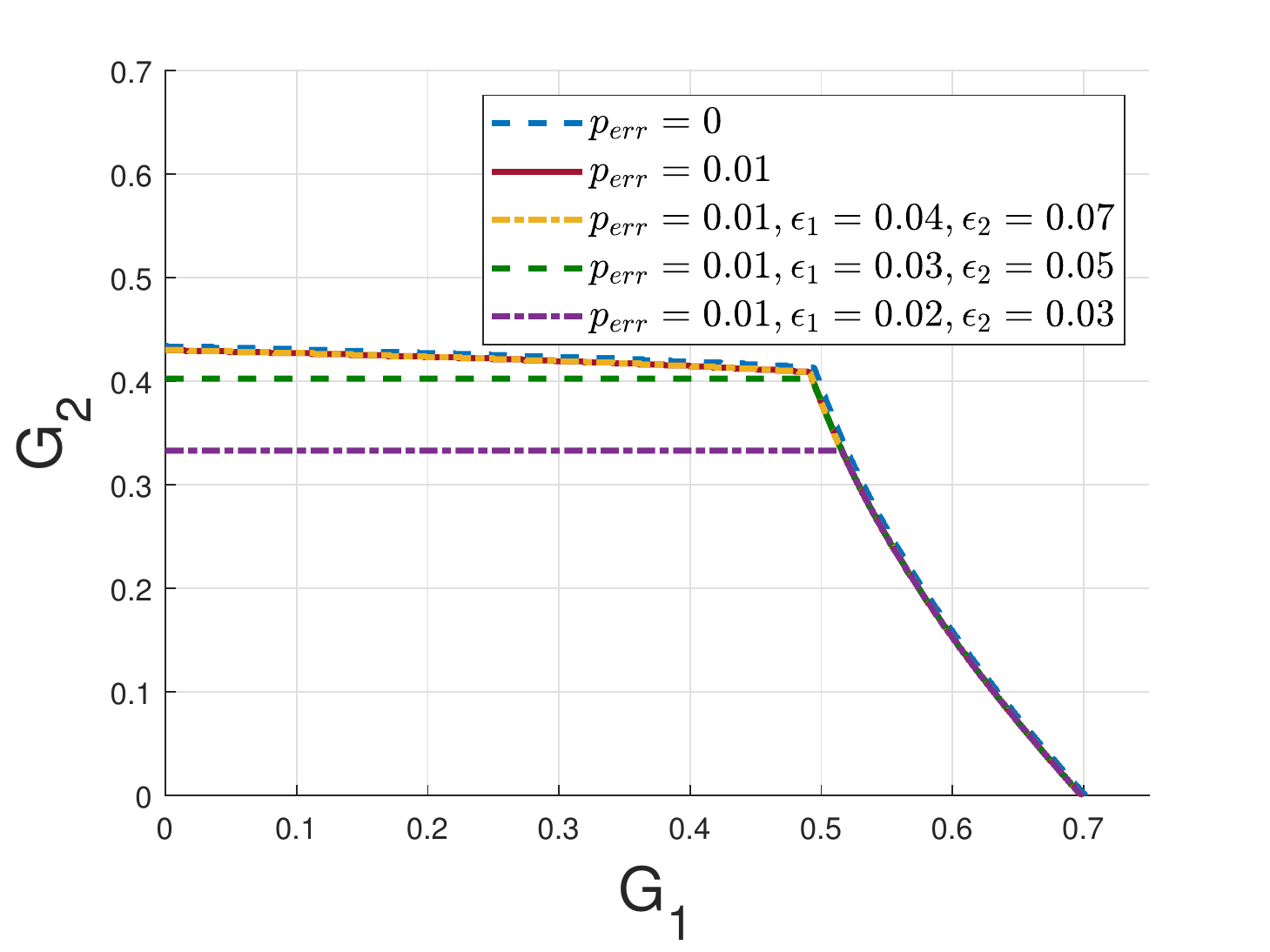}\\
			(a) Complete sharing. & (b) Receiver reservation.
		\end{tabular}
		\caption{The stability region, the weak stability region, and the $\epsilon$-stability region for the complete sharing policy and the receiver reservation policy.}
		\label{fig:weakstability}
	\end{center}
\end{figure}

In \rfig{weakstability}, we show the stability region, the weak stability region, and the $\epsilon$-stability region for the complete sharing policy and the receiver reservation policy. As shown in \rfig{weakstability}(a), the weak stability region for $p_{\rm err}=0.01$ is almost the same as the stability region for no decoding errors, i.e., $p_{\rm err}=0$. Moreover, if we select $\epsilon_1$ and $\epsilon_2$ to be 0.07, then the $\epsilon$-stability region is almost identical to the weak stability region. However, if we select a smaller  $\epsilon_1=\epsilon_2=0.04$ (resp. $0.02$), then we can see that the $\epsilon$-stability region is getting smaller (see the green (resp. purple) curve in \rfig{weakstability}(a)).
Since a small $\epsilon$ implies a larger probability for a packet to be successfully received (\rthe{wstable}),
the numerical results in  \rfig{weakstability}(a) show the trade-off between the admissible offered load and the QoS constraint for the packet success probability. One interesting finding of this numerical example is that
 decoding errors affect class 2 users more than class 1 users. This is because we choose the optimized degree distribution for class 2 users from Table 1  of \cite{liva2011graph}, and such an optimized distribution is not as fault tolerant as the simple repetition code for class 1 users.
Similar findings for the reservation policy are shown in  \rfig{weakstability}(b).

\bsubsec{Rayleigh  block fading channel}{rayleigh}

\begin{figure}[ht]
	\centering
	\includegraphics[width=0.43\textwidth]{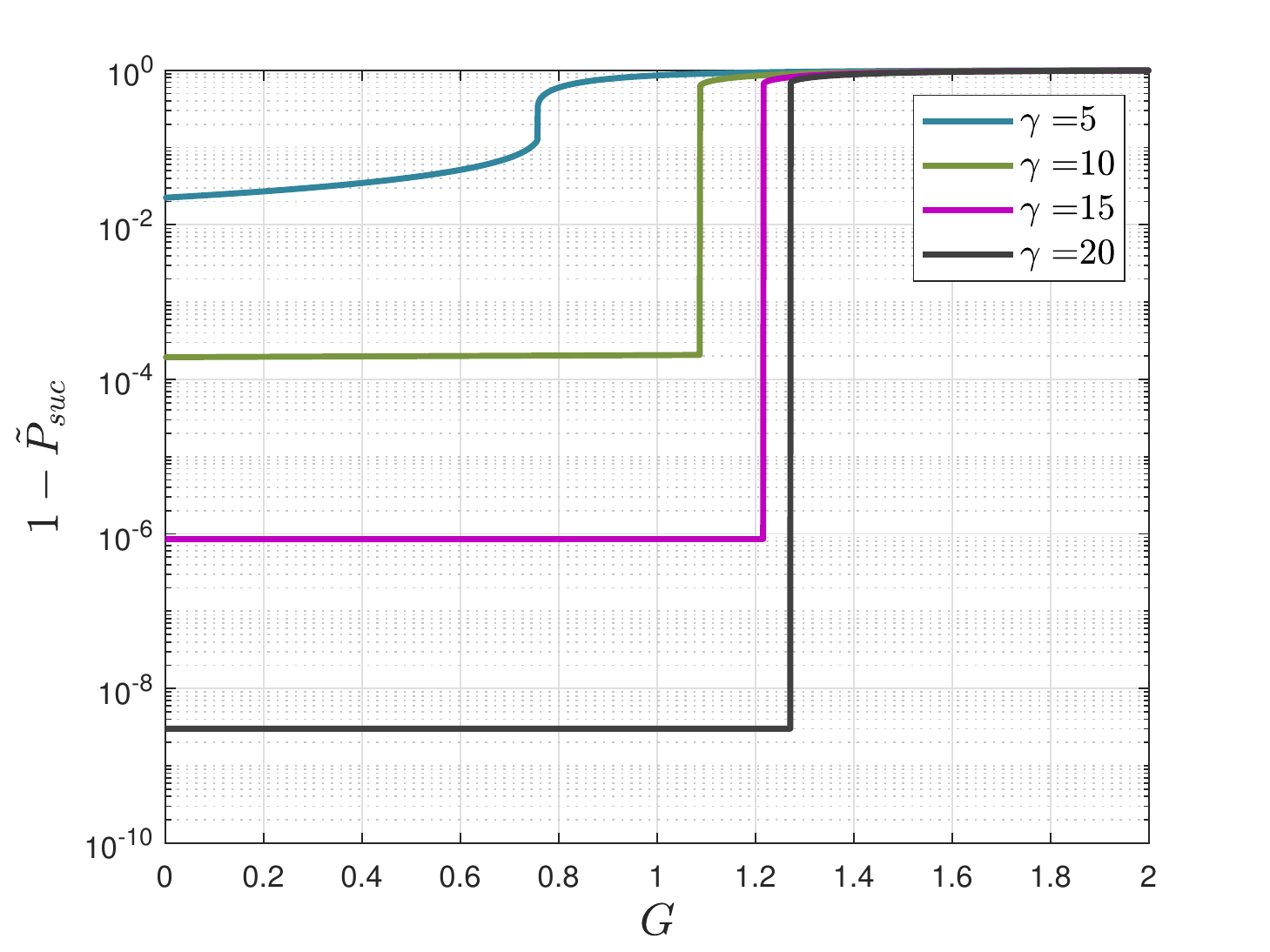}
	\caption{The complement of the success probability function for the Rayleigh  block fading channel  with $\gamma=5,10,15,20dB$.}
	\label{fig:ray}
\end{figure}

\begin{figure}[tb]
	\begin{center}
		\begin{tabular}{p{0.23\textwidth}p{0.23\textwidth}}
			\includegraphics[width=0.23\textwidth]{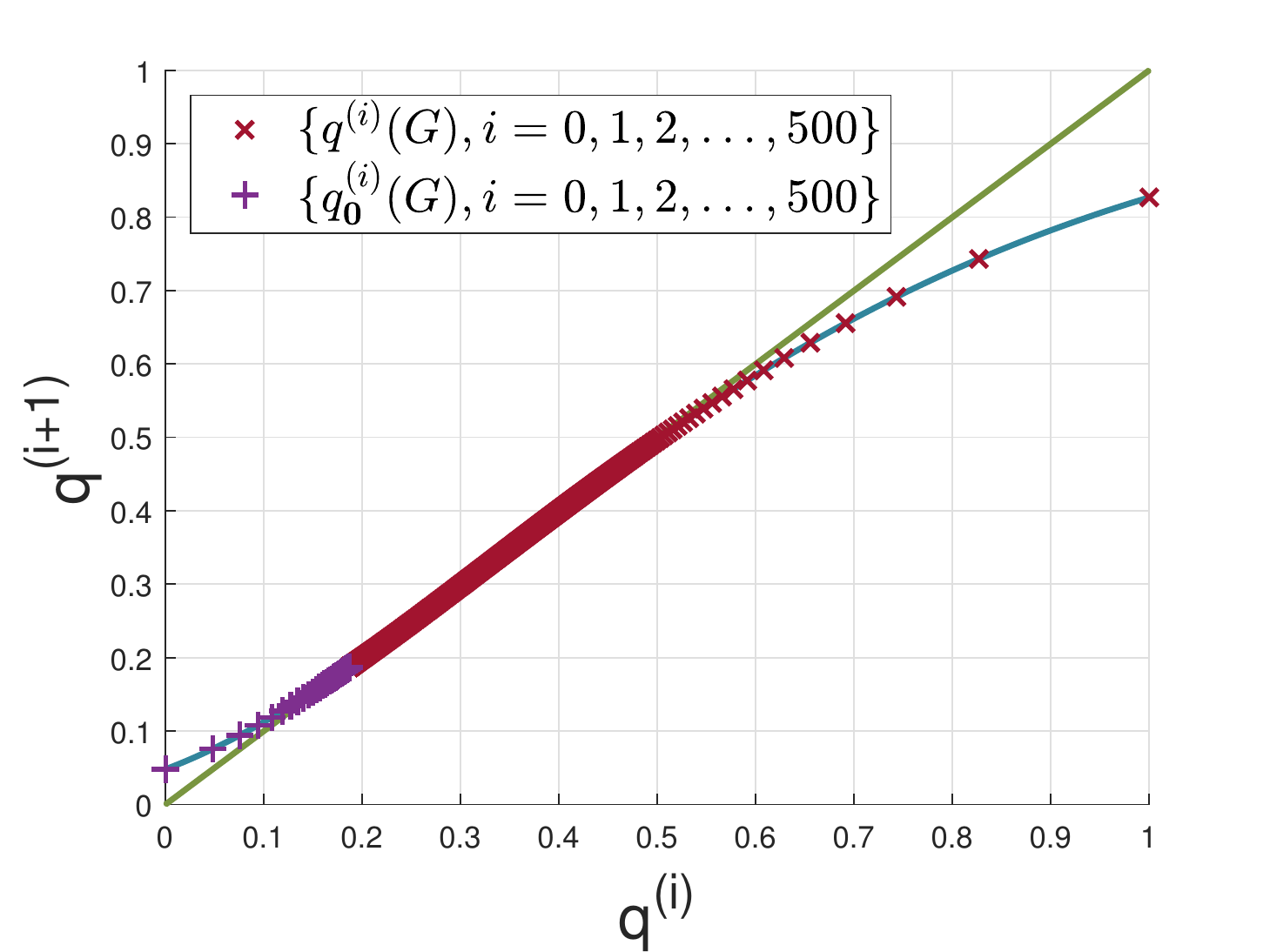} &
			\includegraphics[width=0.23\textwidth]{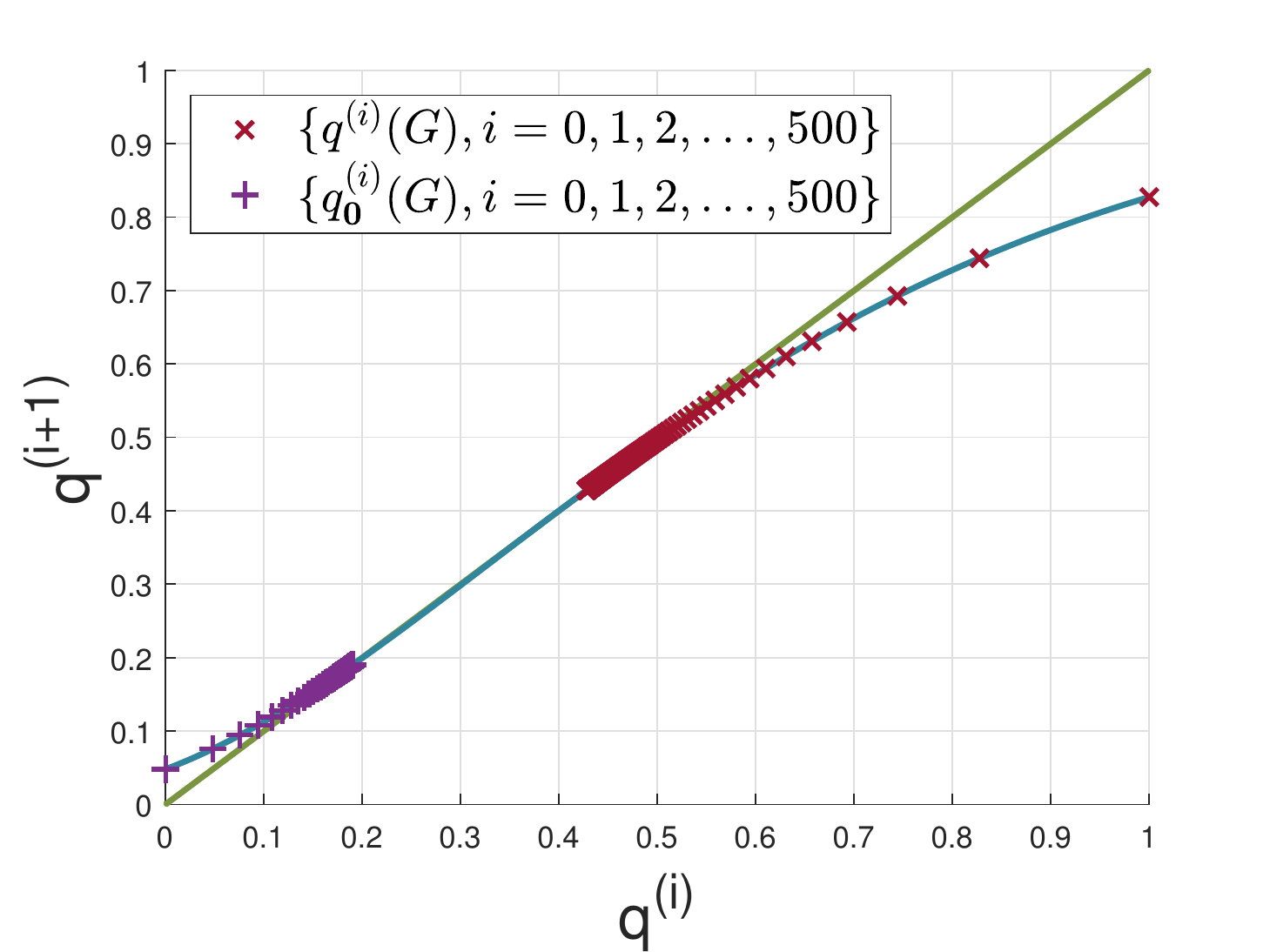}\\
			(a) $G =0.756$. & (b) $G=0.757$.
		\end{tabular}
		\caption{The transfer function from $q^{(i)}$ to $q^{(i+1)}$ in \req{tag6666dmulm} for the Rayleigh  block fading channel  with $\gamma=5dB$.}
		\label{fig:transfer}
	\end{center}
\end{figure}

In this section, we consider a system of coded Poisson receivers with each Poisson receiver modelled by the Poisson receiver for the Rayleigh  block fading channel with capture in \rex{Rayleigh}.
We only consider a single class of users ($K=1$) and a single class of receivers ($J=1$). In such a CPR system, there are $GT$ users with the degree distribution $\Lambda(x)=x^5$ and $T$ independent Poisson receivers with the success probability function in \req{fading8888r}.
By setting  $\gamma=5,10,15,20 dB$ and $b=3 dB$,
 we use \req{mean8888thumulam} to plot the complement of the success probability function $1-\tilde P^{(i)}_{\rm suc}(G)$  for $i=500$ iterations in \rfig{ray}. It is clear that the success probability function in this figure is still of threshold type.  The thresholds $G^*$  are roughly $0.756,1.086,1.215,1.270$ for $\gamma=5,10,15,20dB$, respectively. However, the complement of the success probability function $\tilde P^{(\infty)}(G)$ is not equal to 0 for $G< G^*$.
As such, the notion of stability can no longer be applied, and we have to resort to the weak stability and the  $\epsilon$-stability. As there is only a single class of users, the weak stability region is the region below the threshold $G^*$. To see this, we set $\gamma=5dB$ and use \req{tag6666dmulm} to plot in \rfig{transfer} the transfer functions from $q^{(i)}$ to $q^{(i+1)}$ for  $G=0.756$ and $G=0.757$, respectively. For this setting, the threshold $G^*$ is roughly 0.756. The sequence $\{q^{(i)}(G), i=0,1,2,\ldots\}$ (resp.
 $\{q_{\bf 0}^{(i)}(G), i=0,1,2,\ldots\}$ are the data points  marked with a ``$\times$'' (resp. a ``$+$''). As shown in \rfig{transfer}(a), these two sequences converge to the same solution for $G=0.756$. However, they converge to two different solutions for $G=0.757$.

Now we discuss the $\epsilon$-stability region.
For $\gamma=20dB$ and $b=3dB$,
we have from \req{fading8888r} that
$$\lim_{\rho \to 0}P_{\rm suc}(\rho)=e^{-b/\gamma} \approx 0.980245.$$
To meet the assumption in (A1g), we may choose
$$\epsilon=1-0.980245=0.019755.$$
As a direct consequence of \rthe{wstable}, we have
\beq{wsrates1111r}
1- \tilde P_{{\rm suc}}^{(\infty)}(G)\le \Lambda_k (\epsilon) = \epsilon^5 \approx 3.00866\times 10^{-9},
\eeq
which matches extremely well with the numerical results in \rfig{ray}. For such an $\epsilon$, the $\epsilon$-stability region is almost the same as the weak stability region, and we can admit the load until $G^*$ without sacrificing  the QoS requirement for the packet success probability.
Similar results also hold for  $\gamma=10$ and $15 dB$ where $\lim_{\rho \to 0}P_{\rm suc}(\rho)\approx 1$. However, for the case $\gamma=5 dB$,
$\lim_{\rho \to 0}P_{\rm suc}(\rho)\approx 0.532082$, and it is  far away from 1.
In this case, the curve below the threshold 0.756 is not as flat as the other three cases.
For this case, there is a clear trade-off between the offered load and the QoS constraint for the packet success probability.

\subsection{Codes Poisson receivers with two classes of users and one class of receivers}
\label{sec:coop}

In this section, we consider a system of coded Poisson receivers with each Poisson receiver modelled by the Poisson receiver in \rex{tworeceiversb}. For this system of coded Poisson receivers,  we consider the setting with $G_3T$ URLLC (class 3) users and $G_4T$ eMBB (class 4) users.
We only consider a single class of receivers ($J=1$) for this system of coded Poisson receivers.

Our objective in this section is to show the effect of degree distributions on the stability region.
Let $\Lambda_3(x)$ (resp. $\Lambda_4(x)$) be the degree distribution of class 3 (resp. 4) users.
We consider the following five pairs of degree distributions: (i) $\Lambda_3(x)=\Lambda_4(x)=x^2$, (ii) $\Lambda_3(x)=\Lambda_4(x)=x^3$, (iii) $\Lambda_3(x)=\Lambda_4(x)=x^4$, (iv) $\Lambda_3(x)=\Lambda_4(x)=0.5102x^2+0.4898x^4$, and (v) $\Lambda_3(x)=\Lambda_4(x)=0.5x^2+0.28x^3+0.22x^8$.
We note that the degree distributions in (iv) and (v) are selected  from \cite{liva2011graph} for IRSA with a single class of users. Also, if $G_4=0$, then the system is reduced to
IRSA with $G_3T$ class 3 users in $T$ time slots. As shown in \cite{liva2011graph}, the stability region for the degree distribution in (iv) (resp. (v)) is $G_3 \le 0.868$ (resp. $G_3 \le 0.938$) when $G_4=0$.
On the other hand, if $G_3=0$, then the system is reduced to IRSA with $G_4T$ class 4 users in $2T$ time slots.
Thus, the stability region for the degree distribution in (iv) (resp. (v)) is $G_4 \le 0.868*2$ (resp. $G_4 \le 0.938*2$) when $G_3=0$.
To identify the general stability regions for these five pairs of degree distributions,
we use \req{mean8888thumulam} in \rthe{mainext} to compute the success probabilities $\tilde P_{{\rm suc,k}}^{(i)}(G_3,G_4)$, $k=3$ and 4, after 500 SIC iterations ($i=500$)
for each pair of $(G_3, G_4)$ with the step size of 0.001.

\begin{figure}[ht]
	\centering
	\includegraphics[width=0.43\textwidth]{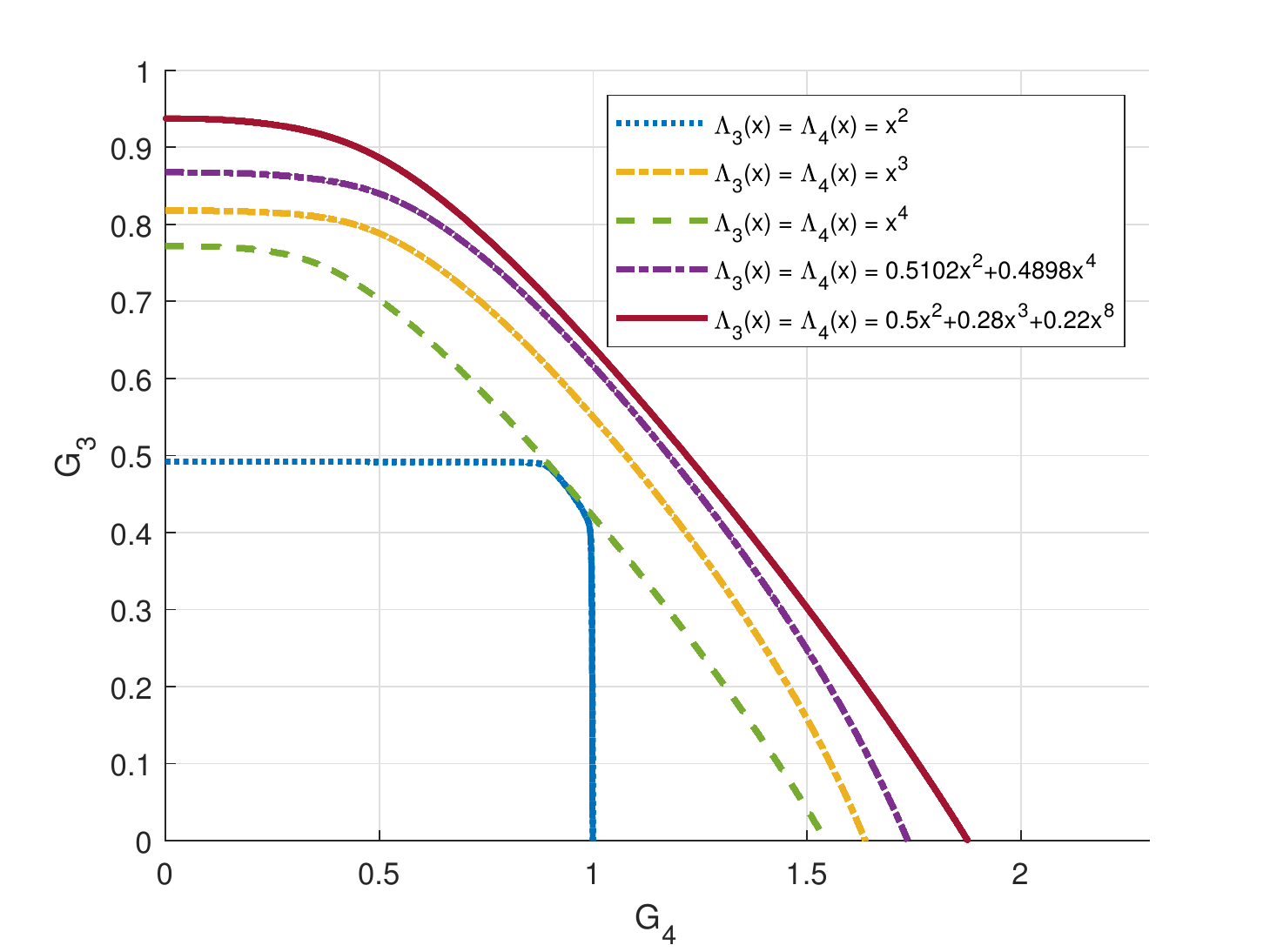}
	\caption{The effect of the degree distributions on the stability region.}
	\label{fig:sd}
\end{figure}

In \rfig{sd}, we show the stability regions for these five pairs of degree distributions.
Clearly, the degree distributions in (v) achieve the largest stability region. However, this is at the cost of a large degree (the degree is 8). This means that a packet might be transmitted 8 times and that consumes 8 times of energy. It seems that the degree distributions in (iv) is a reasonably good choice if we would like to keep the maximum degree under 4. However, as we commented in Section \ref{sec:weaknum}, such an optimized degree distribution might be more susceptible to errors than the simple repletion code in (ii).

\section{Conclusion}
\label{sec:con}

In this paper, we
 extended the probabilistic analysis of CPR systems to the setting with multiple classes of users and receivers.
For such a CPR system, we prove under (A1)-(A3)  that there is a stability region such that each transmitted packet can be successfully received with probability 1 when the offered load to the system is within the stability region. On the other hand, if the offered load is outside the stability region, there is a nonzero probability that a packet will fail  to be received. To cope with noise, decoding errors, and channel fading, we further extended the notion of stability region to weak stability region and $\epsilon$-stability region.
We also demonstrated the capability of providing  differentiated QoS in such CPR systems by comparing the stability regions under various packet routing policies.

There are several possible extensions of this paper:

\noindent (i) {\em Connections to the ALOHA receivers in \cite{liu2020aloha}:} a $\phi$-ALOHA receiver in \cite{liu2020aloha} specifies a deterministic function $\phi$ that maps the number of arriving packets to the number of successfully received packets. As $T \to \infty$,  we have from the law of large numbers that the fraction of the number of successfully received packets to the number of arriving packets in a CPR system becomes a deterministic function of the load vector $G$. In other words, a CPR system can be viewed as a $\phi$-ALOHA receiver in \cite{liu2020aloha} and the network calculus developed in \cite{liu2020aloha} might be applicable for analyzing more complicated CPR systems than those considered in this paper.

\noindent (ii) {\em Optimizing the system parameters:} in this paper, we only show how the routing probabilities and the degree  distributions affect the stability regions. A more general approach is to consider utility functions (see, e.g., \cite{anand2020joint}) and optimize the routing probabilities by solving constrained optimization problems.

\noindent (iii) {\em The condition for the existence of the weak stability region:} the notion of weak stability is defined by the unique solution of a set of equations. Identifying a necessary and sufficient  condition appears to be difficult as it might involve the higher order properties of the success probability functions.

\begin{IEEEbiography}[{\includegraphics[width=1in,height=1.25in,clip,keepaspectratio]{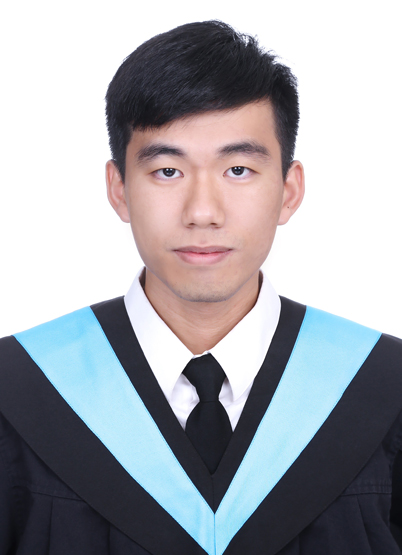}}]
{Chia-Ming Chang} received the B.S. degree in electrical and computer engineering from National Chiao Tung University, Hsinchu, Taiwan (R.O.C.), in 2019. He is currently pursuing the M.S. degree in the Institute of Communications Engineering, National Tsing Hua University, Hsinchu, Taiwan (R.O.C.). His research interest is in 5G wireless communication.
\end{IEEEbiography}

\begin{IEEEbiography}[{\includegraphics[width=1in,height=1.25in,clip,keepaspectratio]{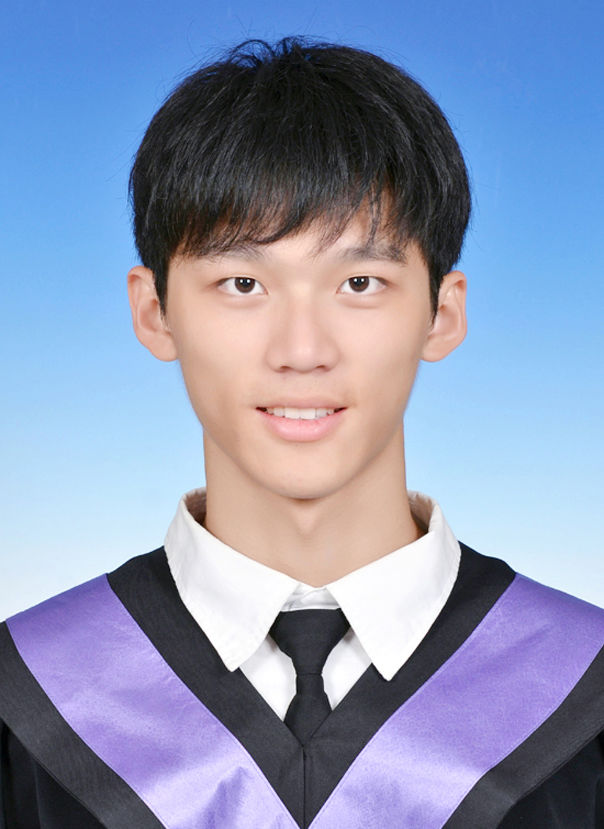}}]
{Yi-Jheng Lin} received his B.S. degree in electrical engineering from National Tsing Hua University, Hsinchu, Taiwan, in 2018. He is currently pursuing the Ph.D. degree in the Institute of Communications Engineering, National Tsing Hua University, Hsinchu, Taiwan. His research interests include wireless communication and cognitive radio networks.
\end{IEEEbiography}

\begin{IEEEbiography}[{\includegraphics[width=1in,height=1.25in,clip,keepaspectratio]{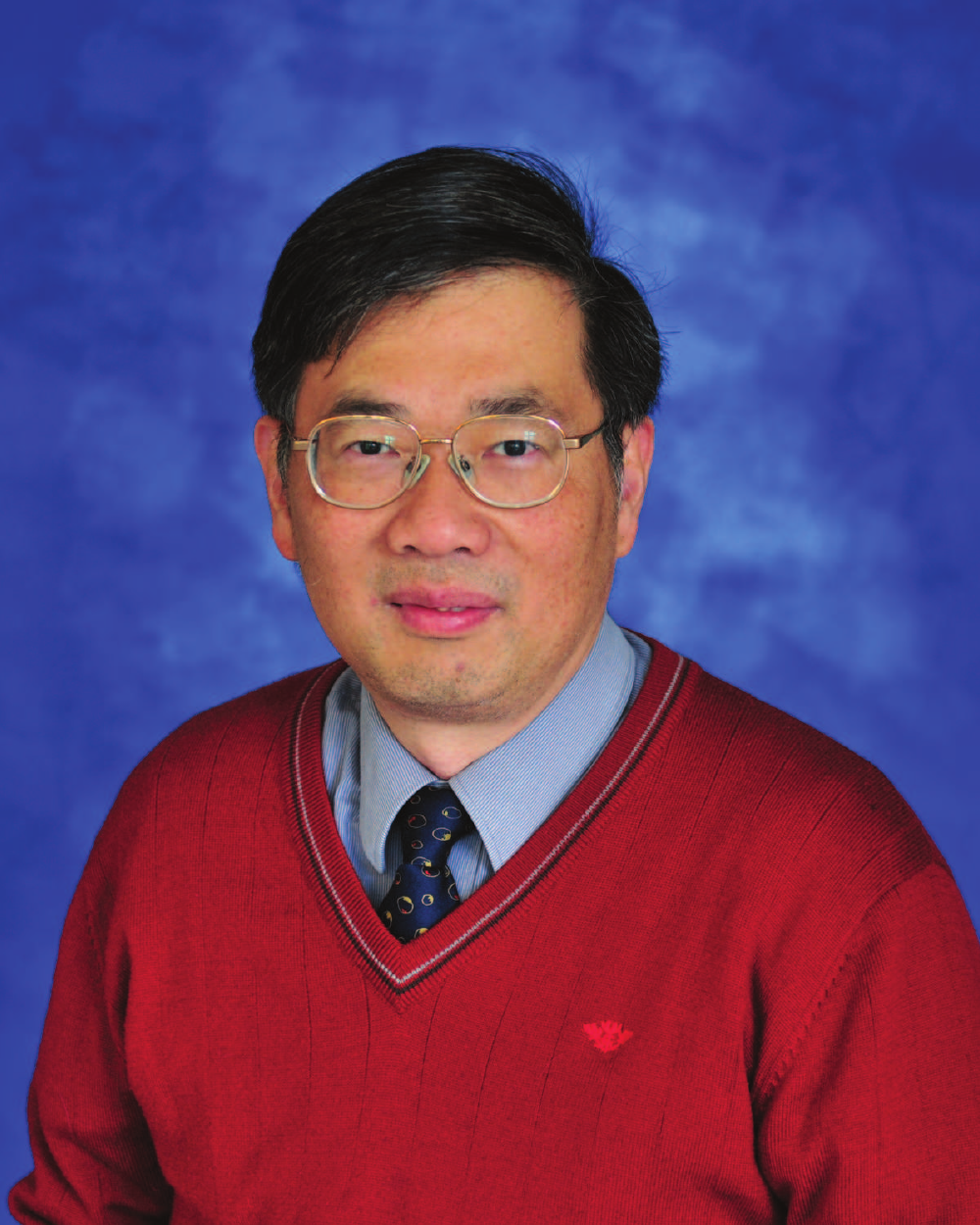}}]
	{Cheng-Shang Chang}
	(S'85-M'86-M'89-SM'93-F'04)
	received the B.S. degree from National Taiwan
	University, Taipei, Taiwan, in 1983, and the M.S.
	and Ph.D. degrees from Columbia University, New
	York, NY, USA, in 1986 and 1989, respectively, all
	in electrical engineering.
	
	From 1989 to 1993, he was employed as a
	Research Staff Member with the IBM Thomas J.
	Watson Research Center, Yorktown Heights, NY,
	USA. Since 1993, he has been with the Department
	of Electrical Engineering, National Tsing Hua
	University, Taiwan, where he is a Tsing Hua Distinguished Chair Professor. He is the author
	of the book Performance Guarantees in Communication Networks (Springer,
	2000) and the coauthor of the book Principles, Architectures and Mathematical
	Theory of High Performance Packet Switches (Ministry of Education, R.O.C.,
	2006). His current research interests are concerned with network science, big data analytics,
	mathematical modeling of the Internet, and high-speed switching.
	
	Dr. Chang served as an Editor for Operations Research from 1992 to 1999,
	an Editor for the {\em IEEE/ACM TRANSACTIONS ON NETWORKING} from 2007
	to 2009, and an Editor for the {\em IEEE TRANSACTIONS
		ON NETWORK SCIENCE AND ENGINEERING} from 2014 to 2017. He is currently serving as an Editor-at-Large for the {\em IEEE/ACM
		TRANSACTIONS ON NETWORKING}. He is a member of IFIP Working
	Group 7.3. He received an IBM Outstanding Innovation Award in 1992, an
	IBM Faculty Partnership Award in 2001, and Outstanding Research Awards
	from the National Science Council, Taiwan, in 1998, 2000, and 2002, respectively.
	He also received Outstanding Teaching Awards from both the College
	of EECS and the university itself in 2003. He was appointed as the first Y. Z.
	Hsu Scientific Chair Professor in 2002. He received the Merit NSC Research Fellow Award from the
	National Science Council, R.O.C. in 2011. He also received the Academic Award in 2011 and the National Chair Professorship in 2017 from
	the Ministry of Education, R.O.C. He is the recipient of the 2017 IEEE INFOCOM Achievement Award.
\end{IEEEbiography}

\begin{IEEEbiography}
	[{\includegraphics[width=1in,height=1.25in,clip,keepaspectratio]{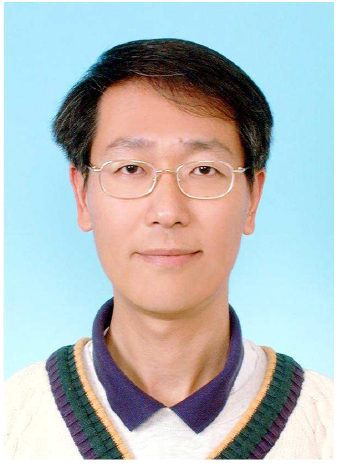}}]
	{Duan-Shin Lee}(S'89-M'90-SM'98) received the B.S. degree from National Tsing Hua
	University, Taiwan, in 1983, and the MS and Ph.D. degrees from
	Columbia University, New York, in 1987 and 1990, all in electrical
	engineering.  He worked as a research staff member at the C\&C Research Laboratory
	of NEC USA, Inc. in Princeton, New Jersey from 1990 to 1998.  He joined the
	Department of Computer Science of National Tsing Hua University in Hsinchu,
	Taiwan, in 1998.  Since August 2003, he has been a professor.  He received
	a best paper award from the Y.Z. Hsu Foundation in 2006.  He served as
	an editor for the Journal of Information Science and Engineering between
	2013 and 2015.  He is currently an editor for Performance Evaluation.
	Dr. Lee's current research interests are network science, game theory,
	machine learning and high-speed networks.  He is a senior IEEE member.
\end{IEEEbiography}


\appendix
\section*{Appendix A}

\setcounter{section}{1}

In this section, we provide a list of notations used in this paper.

{
	\tiny
	\setlength{\tabcolsep}{10pt}
	\renewcommand\arraystretch{0.4}
	\begin{table}[ht]
			\caption{List of Notations\label{tab:one}}
			{%
				\scalebox{0.9}{
				\begin{tabular}{||l|l||}
					
				\hline\hline

				{\color{black}$b$} & The threshold for the Rayleigh block fading channel\\
				{\color{black}$D$} & The maximum number of packets that can be successfully \\
				& received in $D$-fold ALOHA\\
				{\color{black}$F_j$} & The fraction of Poisson receivers assigned to class $j$ receivers \\
				$G$ & {\color{black}A vector of normalized offered load, }\\
				& {\color{black} $G=(G_1, G_2, \ldots, G_{K})$ }\\
				{\color{black}$G_k$} & The normalized offered load of class $k$\\
				{\color{black}$G^*$} & The percolation threshold of $G$\\
				{\color{black}$i$} & The number of iteration\\
				$J$ & The number of classes of receivers \\
				$K$ & The number of classes of users \\
				$L_k$ & The number of copies of a class $k$ packet \\
				{\color{black}$N$} & The number of active users in the Rayleigh \\ & block fading
channel\\
				$P_{\rm suc}$ & The success probability of a tagged packet \\
				{\color{black}$\tilde P_{{\rm suc},k}^{(i)}$} &
				The probability that a packet sent from a randomly selected \\
				& {\em class $k$ user} can be successfully received after the $i^{th}$ iteration\\
				{\color{black}$p_{\rm era}$} & The probability that a packet is erased (due to fading)\\
				{\color{black}$p_{\rm err}$} & The probability of decoding errors in a Poisson receiver\\
				{\color{black}$p_{\rm sic}$} & The probability that a copy of a successfully received packet \\
				& cannot be removed from the SIC operation\\
				$p_k^{(i)}$ & The probability that the receiver end of a randomly \\
				& selected class $k$ edge has not been successfully\\
				& received after the $i^{th}$ SIC iteration \\
				{\color{black}$p_{k,j}^{(i)}$} & The probability that the receiver end of a randomly \\
				&selected class $(k,j)$-edge has not been successfully\\
				& received after the $i^{th}$ SIC iteration\\
				{\color{black}$q^{(i)}$} & A vector of $q_k^{(i)}$, $q^{(i)}=(q_{1}^{(i)}, q_{2}^{(i)}, \ldots, q_{K}^{(i)})$\\
				{\color{black}$q^{(0)}$} & The initial vector of $q^{(i)}$ \\
				$q_k^{(i)}$ & The probability that the user end of a randomly \\
				& selected class $k$ edge has not been successfully \\
				& received after the $i^{th}$ SIC iteration\\
				{\color{black}$q^{(\infty)}$} & The limiting vector of $q^{(i)}$ with the initial vector $q^{(0)}={\bf 1}$ \\
				{\color{black}$\qinf$} & The limiting vector of $q^{(i)}$ with the initial vector $q^{(0)}={\bf 0}$ \\
				{\color{black}$\Rj$} & A vector of parameters for class $j$ receiver, \\
				& $\Rj=(\frac{r_{1,j}}{F_j}, \frac{r_{2,j}}{F_j}, \ldots, \frac{r_{K,j}}{F_j})$\\
				$r_{k,j}$ & The routing probability that a class $k$ packet \\
				&transmitted to a class {\color{black}$j$} receiver \\
				$S$ & {\color{black} The stability region}\\
				{\color{black}$S(\epsilon)$} & The $\epsilon$-stability region\\
				$T$ & The number of Poisson receivers\\
				{\color{black}$X$} & The exponentially distributed random variable with mean $1$ \\
				{\color{black}$\Gamma_0(\epsilon)$} & A nonempty set for $\rho$, $\rho \in \Gamma_0(\epsilon)$ when
				$P_{{\rm suc},k,j}(\rho)\ge 1-\epsilon_k$ \\
				& for $j=1,2, \ldots, J$, and $k=1,2, \ldots, K$\\
				{\color{black}$\Gamma_1(\epsilon)$} & A bounded region of $q$, \\
				& $\Gamma_1(\epsilon)=[0, \lambda_1(\epsilon_1)] \times [0, \lambda_2(\epsilon_2)] \times \ldots \times  [0, \lambda_K(\epsilon_K)]$\\
				{\color{black}$\Gamma_2(\epsilon)$} & A bounded region of $p$, \\
				& $\Gamma_2(\epsilon)=[0, \epsilon_1] \times [0,\epsilon_2]\times \ldots \times [0,\epsilon_K]$\\
				{\color{black}$\gamma$} & The signal-to-noise ratio\\
				$\epsilon$ & A vector of parameters for $\epsilon$-stability, {\color{black}$\epsilon=(\epsilon_1,\epsilon_2,\ldots, \epsilon_K)$}\\
				{\color{black}$\Theta_k$} & The throughput of class $k$ packets\\               				
				$\Lambda_{k,\ell}$ & The probability that a class $k$ packet is transmitted\\
				& $\ell$ times \\
				$\Lambda_k(x)$ & The generating function of the degree distribution of\\
				& a class $k$ user \\
				{\color{black}$\Lambda_{k}^\prime(x)$} & The derivative of $\Lambda_k(x)$\\
				{\color{black}$\Lambda^\prime(x)$} &
				A vector of $\Lambda_{k}^\prime(x)$, $\Lambda^\prime (x)=(\Lambda^\prime_1 (x), \Lambda^\prime_2 (x), \ldots, \Lambda^\prime_K (x))$\\
				{\color{black}$\Lambda_{k}^\prime(1)$} & The mean degree of a class $k$
				user node\\
				$\lambda_{k,\ell}$ & The probability that the user end of a randomly \\
				&selected class $k$ edge has additional $\ell$ edges\\
				&excluding the randomly selected  edge \\
				$\lambda_k(x)$ & The generating function of the excess degree \\
				&distribution of a class $k$ user \\
				{\color{black}$\lambda_k^{-1}(\cdot)$} & The inverse function of $\lambda_k(x)$ \\
				$\rho$ & The Poisson offered load $\rho=(\rho_1, \ldots, \rho_K)$\\
				$\rho_k$ & The Poisson offered load of class $k$\\
				{\color{black}$\rho_{k,j}$} & The Poisson offered load of class $k$ packets to a class $j$ \\
				& Poisson receiver\\
				{\color{black}$\trhoj$} & The Poisson offered load at a class $j$ Poisson receiver\\
				& $\trhoj=(\rho_{1,j}, \rho_{2,j}, \ldots, \rho_{K,j})$\\
				
				\hline
				\hline

			\end{tabular}}}
			\label{table:notations}
	\end{table}
}

\end{document}